\documentclass[runningheads]{llncs}

\newif\ifextended
\extendedtrue

\usepackage[T1]{fontenc}
\usepackage{listings}
\usepackage{graphicx}
\usepackage{xspace}
\usepackage{todonotes}
\usepackage{xargs}
\usepackage{caption}
\usepackage{multirow}
\usepackage{multicol}
\usepackage{lmodern}
\usepackage{pgfplots}
\usepackage{mathtools}
\usepackage{amsmath}
\usepackage{tikz}
\usepackage{caption}
\usepackage{subcaption}
\usepackage[hidelinks]{hyperref}
\usepackage{cleveref}
\usepackage{gobble}
\usepackage{amssymb}
\usepackage{amsfonts}
\usepackage{xspace}
\usepackage{tabularx}
\usepackage{paralist}
\usepackage{ragged2e}
\usepackage{arydshln}
\usepackage[ligature, inference]{semantic}
\usetikzlibrary{automata, positioning, arrows,quotes}
 \usetikzlibrary{calc}
\usepackage[inference]{semantic}

\makeatletter
\newcommand\notsotiny{\@setfontsize\notsotiny{6}{7}}
\newcommand\testfont{\@setfontsize\notsotiny{6}{7}}
\makeatother

\newcommand{\Nat}{\ensuremath{\mathbb{N}}\xspace}
\newcommand{\Tids}{\ensuremath{\mathbb{T}}\xspace}
\newcommand{\ALoc}{\ensuremath{\mathbb{A}}\xspace}

\newcommand{\Reg}{\ensuremath{\textit{Reg}}\xspace}
\newcommand{\Values}{\ensuremath{\mathbb{V}\xspace}}

\newcommand{\Local}{\ensuremath{\mathcal{L}\xspace}}

\newcommand{\Read}{\ensuremath{\mathsf{Read}}\xspace}
\newcommand{\PRead}{\ensuremath{\mathsf{PRead}}\xspace}
\newcommand{\Latest}{\ensuremath{\pi}\xspace}
\newcommand{\Time}{\ensuremath{\tau}\xspace}
\newcommand{\CASsuccess}{\ensuremath{\sigma}\xspace}

\newcommand{\wrlx}{\ensuremath{w\!-\!rlx}\xspace}
\newcommand{\rrlx}{\ensuremath{r\!-\!rlx}\xspace}
\newcommand{\rrlxs}{\ensuremath{r\!-\!rlx^*}\xspace}

\newcommand{\Views}{\ensuremath{\mathbb{W}}\xspace}
\newcommand{\States}{\ensuremath{\mathbb{S}}\xspace}
\newcommand{\InitStates}{\ensuremath{\hat{\mathbb{S}}}\xspace}

\newcommand{\getname}{\texttt{load}\xspace}
\newcommand{\setname}{\texttt{store}\xspace}
\newcommand{\casname}{\texttt{rmw}\xspace}

\newcommand{\txrightarrow}{\ensuremath{\rightarrow_{t,x}}\xspace}

\newcommand{\newparagraph}[1]{\vspace{0.1cm} \noindent\textit{#1}}
\newcommand{\stt}[3]{\ensuremath{s^{#2,#3}_{#1}}}
\newcommand{\stti}[3]{\ensuremath{\hat{s}^{#2,#3}_{#1}}}

\definecolor{ao}{rgb}{0.0, 0.3, 0.0}
\definecolor{burgundy}{rgb}{0.5, 0.0, 0.13}
\definecolor{ceruleanblue}{rgb}{0.10, 0.30, 0.90}

\lstdefinestyle{myinlinestyle}{
	float=t,
	numbers=none,
	columns=flexible,
	commentstyle=\color{black},
	framexleftmargin=1mm, 
	xleftmargin=1mm, 
	alsoletter=\,*,
	breaklines=true,
	tabsize=2,
	basicstyle=\sf,
	framesep=\fboxsep,
	framerule=\fboxrule,
	mathescape=true,
	showstringspaces=false,
	showspaces=false,
	keywordstyle=\sf,
	escapechar={\#}
}

\lstdefinestyle{myCustomCStyle3}{
	float=t,
	language = C,
	numbers=left,
	commentstyle=\color{ceruleanblue},
	columns=flexible,
	xleftmargin=1mm, 
	alsoletter=\,*,
	morekeywords={invariant, ensures, requires, context, context_everywhere, \old, \length, par,
		\forall, \forall*, **, Perm,send,receive,__kernel,loop_invariant,for, Barrier, barrier, resource, CLK_GLOBAL_MEM_FENCE, get_global_id, seq, frac, inline, tail, head, none, }, 
	breaklines=true,
	tabsize=4,
	basicstyle=\ttfamily\scriptsize,
	keywordstyle=\color{burgundy}\tt,
	frame = single,
	framesep=\fboxsep,
	framerule=\fboxrule,
	mathescape=true,
	showstringspaces=false,
	showspaces=false,
	escapechar={\#},
	xleftmargin=0.05\textwidth,
	xrightmargin=0.01\textwidth
}

\lstdefinestyle{myCustomCStyle2}{
	float=t,
	language = C,
	commentstyle=\color{ao},
	columns=flexible,
	framexleftmargin=1mm, 
	numbers=left,
	stepnumber=1,
	xleftmargin=1mm, 
	morekeywords={pragma,omp,nowait,schedule, shared,parallel,section,simd,single,sections,For,WhileVec,Par,ParVec,encode,compose, translate, fusiable, par_able,sec,m,bundle},
	breaklines=true,
	numberstyle=\tiny\sf,
	tabsize=2,
	basicstyle=\footnotesize\tt,
	keywordstyle=\color{blue}\tt,
	framesep=\fboxsep,
	framerule=\fboxrule,
	mathescape=true,
	showstringspaces=false,
	showspaces=false,
	escapechar={\$}
}

\newcommandx{\yaHelper}[2][1=\empty]{%
	\ifthenelse{\equal{#1}{\empty}}%
	{ \ensuremath{ \scriptstyle{#2} } } 
	{ \raisebox{#1}[2pt][0pt]{ \ensuremath{ \scriptstyle{#2} } } }  
}

\newcommandx{\yhookrightarrow}[4][1=\empty, 2=\empty, 4=\empty, usedefault=@]{%
	\ifthenelse{\equal{#2}{\empty}}
	{ \xhookrightarrow{ \protect{ \yaHelper[ #4 ]{\!\!#3\!\!} } } } 
	{ \xhookrightarrow[ \protect{ \yaHelper[ #2 ]{ #1 } } ]{ \protect{ \yaHelper[ #4 ]{ #3 } } } }
}

\newcommand{\tightoverset}[2]{\mathop{#2}\limits^{\vbox to -.5ex{\kern-0.75ex\hbox{$#1$}\vss}}}

\newcommand{\code}[1]{\lstinline!#1!}

\crefformat{line}{l.#2#1#3}
\crefrangelabelformat{line}{l.#3#1#4--#5#2#6}
\crefname{line}{}{}
\crefname{figure}{Fig.}{Figs.}
\crefname{section}{Section}{Sections}
\crefname{definition}{Def.}{Defs.}
\crefname{appendix}{Appendix}{Appendices}

\newcommand{\frl}{\color{ceruleanblue}$\forall$}

\DeclareSymbolFont{matha}{OML}{txmi}{m}{it}
\DeclareMathSymbol{\varv}{\mathord}{matha}{118}
\usepackage{scalerel,stackengine,amsmath}
\newcommand\equalhat{\mathrel{\stackon[1.5pt]{=}{\stretchto{%
                \scalerel*[\widthof{=}]{\wedge}{\rule{1ex}{3ex}}}{0.5ex}}}}

\begin{document}
\ifextended
\title{Deductive Verification of Weak Memory Programs with View-based Protocols\texorpdfstring{\\}{\ }(extended version)\thanks{This is a preprint. The Version of Record is published in Model Checking Software (SPIN 2026) and is available at TBD.}}
\titlerunning{Deductive Verification of Weak Memory Programs (extended version)}
\fi
\titlerunning{Deductive Verification of Weak Memory Programs}

\author{
    \"Omer {\c{S}akar}\inst{1}\orcidID{0000-0003-3457-5446} \and
	Soham Chakraborty\inst{2}\orcidID{0000-0002-4454-2050} \and
	Marieke {Huisman}\inst{1}\orcidID{0000-0003-4467-072X} \and
	Anton {Wijs}\inst{3}\orcidID{0000-0002-2071-9624} 
}
\authorrunning{
    \"O. \c{S}akar et al.
    }

\institute{
	University of Twente, The Netherlands \\
	\email{\{o.f.o.sakar, m.huisman\}@utwente.nl} \and
	TU Delft, The Netherlands \\
    \email{s.s.chakraborty@tudelft.nl} \and
	Eindhoven University of Technology, The Netherlands \\
	\email{a.j.wijs@tue.nl}
}

\maketitle              

\begin{abstract}
Concurrent programming under weak memory concurrency faces substantial challenges to ensure correctness due to program behaviors that cannot be explained by thread interleaving, a.k.a. sequential consistency. While several program logics are proposed to reason about weak memory concurrency, their usage has been limited to intricate manual proofs. On the other hand, the VerCors verifier provides a rich toolset for automated deductive verification for sequential consistency.

In this paper, we bridge this gap for automated deductive verification of weak memory concurrent programs with the VerCors deductive verification tool. We propose an approach to encode weak memory concurrency in VerCors. We develop VerCors-relaxed, where we extend the VerCors atomics support and bring concepts from several protocol automata to encode
permission-based separation logics for weak memory concurrency models.
To demonstrate the effectiveness of our approach, we encode the relaxed fragment of the SLR program logic, a recent state-of-the-art permission-based separation logic for weak memory concurrency in VerCors-relaxed, our extension of VerCors. We use the SLR encoding on VerCors-relaxed to automatically verify several examples from the literature within realistic performance. 

\end{abstract}

\newcommand{\splitsecs}{}


\section{Introduction}
\label{sec:intro}

Reasoning about programs with weak memory models is challenging, but necessary, as modern compilers apply a range of optimisations
depending on the targeted hardware. The challenge is that weak memory models allow program executions that are not explainable by a 
sequentially-consistent (SC) interleaving of thread instructions, and may cause program outcomes that are unexpected.

To formally reason about weak memory models, several program logics such as Relaxed Separation Logic (RSL)~\cite{vafeiadis2013relaxed}, GPS~\cite{2014_GPS_turon} and FSL~\cite{doko2015program} have been developed. These logics tackle different (parts of) weak memory models. However, all these logics require proving manually whether or not an execution is allowed by a given weak memory model. 


In this work, we focus on the deductive verification of weak memory programs. 
Previous work in this direction~\cite{Amighi18Chap5,amighi2014resource,2014_GPS_turon} used protocols per location to specify the possible writes to a location in the entire program.
We take a slightly different approach, and use protocols to specify the possible writes per thread-location combination, more specifically, for a thread and a location it intends to write to.
We combine this with thread-local views that capture the progress of all threads. 
These two ingredients are combined into \textit{view-based protocols}.

Concretely, we propose VerCors-relaxed as an approach to encode relaxed memory accesses using view-based protocols in VerCors, a deductive verifier for concurrent programs using
permission-based separation logic (PBSL) with support for SC atomics~\cite{Amighi18Chap5,amighi2014resource}.
The proposed encoding focuses on
usability and automation in modular verification, as it reasons about threads one by one, considering their instructions in SC order.
We demonstrate that our encoding with view-based protocols
provides an elegant way to reason about reordered writes to diﬀerent locations
for all threads in the program, and thus gives the possibility to reason about speculative writes.

To demonstrate the effectiveness of our approach, we encode the relaxed fragment of the SLR program logic~\cite{SvendsenPDLV18}, a recent state-of-the-art logic using PBSL for weak memory concurrency, in VerCors-relaxed. SLR suits our approach well since both use PBSL. We use the SLR encoding in VerCors-relaxed to automatically verify several examples from the literature with realistic performance. 

\newparagraph{Outline \& Contributions.}
\cref{sec:background} introduces a small programming language and VerCors' PVL language.
\cref{sec:20:protocols} explains the first contribution of the paper, namely the concept of view-based protocols and the reasoning with them, using examples.
\cref{sec:encoding_slr} explains SLR, its rules and its encoding using the view-based protocols.
\cref{sec:40:encoding} then explains the second contribution, the encoding of protocols and thread-local views into VerCors.
Several examples from the literature are encoded into VerCors-relaxed, discussed in \cref{sec:evaluation}.
\cref{sec:10:relwork} discusses related work.
Finally, \cref{sec:50:conclu} concludes the paper and discusses future work.


\section{Preliminaries}
\label{sec:background}


\noindent\textbf{Programming language.}
To keep our examples general and not restricted to a particular programming language, we use a basic language to express how registers and shared variables are updated. First of all, a parallel program is executed by a set of threads $\Tids$ that access a set of atomic locations
$\ALoc$ and a set of register variables $\Reg$. Without loss of generality, we assume that all variables in $\ALoc\ \cup \Reg$ are of a generic type $\Values$, and
$v \in \Values$ is a value from this domain.

There are three ways to interact with atomic memory:
\vspace*{5pt}
\begin{compactenum}
    \item Writing of the value $v \in \Values$ to atomic location $x \in \ALoc$ by thread $t \in \Tids$, corresponding to a \emph{relaxed write} instruction, e.g., the \setname operation $x := 1$.  
    \item Reading of the value $v \in \Values$ from atomic location $x \in \ALoc$ by thread $t \in \Tids$, corresponding to a \emph{relaxed read} instruction, e.g., the \getname operation $a := x$ where $a$ is a register/local variable.
    \item A RWM operation of thread $t \in \Tids$ that atomically reads the value $v \in \Values$ from atomic location $x \in \ALoc$ and writes the value $w \in \Values$ to $x$ if $v$ is the expected value, corresponding to a \emph{relaxed read-modify-write}, e.g., \texttt{\casname\!(x,0,2)}.
\end{compactenum}

\vspace*{5pt}\noindent\textbf{VerCors.} 
VerCors is a deductive verifier aimed at concurrent software~\cite{vercorsCAV}. VerCors uses PBSL (Permission-based Separation Logic)~\cite{Berdine05,bornat2005permission,Boyland03}, which allows it to reason about data-race freedom, memory safety, and functional correctness of sequential and concurrent programs. VerCors supports multiple input languages such as Java, C/C++, OpenCL and its language PVL (Prototypal Verification Language), which is mainly used to prototype verification features. It supports classes and methods similar to Java and verification-related concepts. The rest of this section introduces the PVL constructs needed for the work in this paper.

\begin{figure}[t]
	\begin{lstlisting}
class ArraySum {#\label{lst:back:ex:classstart}#
	int sum;#\label{lst:back:ex:fieldsum}#
	/*@#\label{lst:back:ex:contractstart}# context (#\frl# int i=0..N; Perm(xs[i], 1\2)) ** Perm(sum, write);#\label{lst:back:ex:context}#
		requires (#\frl# int i=0..N; 0 <= xs[i]) ** sum == 0;;#\label{lst:back:ex:pre}#
		ensures sum >= 0; #\label{lst:back:ex:post}#@*/#\label{lst:back:ex:contractend}#
	void parSum(int[] xs) {#\label{lst:back:ex:methodstart}#
		invariant Inv(#\label{lst:back:ex:invblockstart}#/*@ #\label{lst:back:ex:invstart}#Perm(sum, write) ** sum >= 0#\label{lst:back:ex:funcprop}# @*/#\label{lst:back:ex:invend}#) {#\label{lst:back:ex:invbodystart}# 
			#\label{lst:back:ex:parstart}#par (int tid=0..N)
			#\label{lst:back:ex:kcontractstart}#/*@ context Perm(xs[tid], 1\2) ** xs[tid] >= 0; @*/#\label{lst:back:ex:kcontractend}#
			{
				atomic(Inv) {#\label{lst:back:ex:atomicstart}#
					sum = sum + xs[tid]; }#\label{lst:back:ex:atomicstop}#}#\label{lst:back:ex:parend}#}#\label{lst:back:ex:invbodyend}##\label{lst:back:ex:invblockend}#}#\label{lst:back:ex:methodend}#}#\label{lst:back:ex:classend}#
\end{lstlisting}
	\caption{A PVL program summing an array in parallel.}
	\label{fig:background:example_pvl}
\end{figure}

An example PVL program is shown in \cref{fig:background:example_pvl}. It uses parallel blocks, invariants and atomics. We use a slightly modified syntax of PVL for readability. In the example, we verify that the sum over positive integers is also positive.

The \texttt{parSum} method in the \texttt{ArraySum} class (\cref{lst:back:ex:methodstart}) has a pre-/postcondition style contract (\crefrange{lst:back:ex:contractstart}{lst:back:ex:contractend}). The keywords \texttt{requires} and \texttt{ensures} specify pre- and postconditions, respectively and \texttt{context} is short for both a pre- and postcondition.

The \texttt{context} statement (\cref{lst:back:ex:context}) specifies permission over the array \texttt{xs}. In PBSL, permissions capture which threads can access which memory locations. To specify permissions, the \texttt{Perm(L, p)} predicate is used where \texttt{L} is a shared heap location and \texttt{p} a fractional in interval \texttt{(0,1]}, where read permission is in \texttt{(0,1)} and write permission is 1 (also written as \texttt{write}). 

Note that $\ast\ast$ and \texttt{\&\&} denote separating (in separation logic) and logical conjunction, respectively and 
that $\forall$* and $\forall$ denote the separating~\cite{MullerSS16} and standard universal conjunction, respectively. Permissions can be split and joined, e.g., \texttt{Perm(L,1$\backslash$2) ** Perm(L,1$\backslash$2)} $\Leftrightarrow$ \texttt{Perm(L,1)}, to distribute permissions over threads. 
VerCors, using the underlying theory, can prove absence of data races (thread safety) by proving that the total permission per memory location does not exceed 1 at any point in time.  

The method contains an \textit{invariant block} (\crefrange{lst:back:ex:invblockstart}{lst:back:ex:invblockend}). The invariant \texttt{Inv} is a statically scoped invariant. It can only be used in the scope of its declaration. It has to hold where it is defined (\cref{lst:back:ex:invstart}) and is used for reasoning about atomic operations (\crefrange{lst:back:ex:atomicstart}{lst:back:ex:atomicstop}), explained below. 
It specifies write permission over the field \texttt{sum} (\cref{lst:back:ex:fieldsum}) and that \texttt{sum} always contains a positive value (including zero). 

The body of the invariant block (\crefrange{lst:back:ex:invbodystart}{lst:back:ex:invbodyend}) contains a parallel block (\cref{lst:back:ex:parstart}), which models the spawning of \texttt{N} threads that execute the body concurrently, with unique identifiers \texttt{tid}. 
The parallel block has a thread contract (\cref{lst:back:ex:kcontractstart}), which specifies the contract for a single thread. 
The thread contract in \cref{fig:background:example_pvl} specifies half (i.e. read) permission  for the location \texttt{xs[tid]} and that \texttt{xs[tid]} is positive.

The parallel block consists of an \textit{atomic block} (\cref{lst:back:ex:atomicstart}), which is executed in one step, i.e., atomically. 
The atomic invariant \texttt{Inv} captures the shared state that is accessed by the atomic instructions, corresponding to the atomic rule in Concurrent Separation Logic~\cite{VAFEIADIS2011335}. The body of the atomic block increments \texttt{sum} by \texttt{xs[tid]}. This is allowed since the invariant \texttt{Inv} contains write permission.


\section{Proposed Approach: Protocols and Thread-local views}
\label{sec:20:protocols}

This section introduces weak memory programs using examples. Next, the concepts of single-thread protocols and thread-local views are explained. The protocols and the thread-local views are used in the encoding of SLR in \cref{sec:encoding_slr}. 
Finally, the verification procedure is explained, by which we reason about coherence properties. 
This procedure is encoded into VerCors-relaxed in \cref{sec:40:encoding}.\\[15pt]
\noindent\textbf{Examples of Weak Memory Programs.}
\cref{fig:intro_example} shows example programs rerevisited throughout this paper.
Register variables are named $a, b, \ldots\in \Reg$, and atomically accessible (shared) locations are named $x, y, \ldots \in \ALoc$.
Threads are labeled $T_0, T_1, \ldots\in\Tids$. All atomic reads and writes use the relaxed memory ordering, on which only
SC-per-location is imposed. We assume that all variables have been initialized to 0. Allowed results are written in blue comments, while
inconsistent results are in red comments. For instance, in the 2+2W (i.e., 2 writes per-thread, 2 in parallel)  example~\cite{kang2017promising}, both $a$ and $b$ may be assigned the value $2$
in the same execution, which is due to the fact that the writes to $x$ and $y$ can be reordered.
However, in the COH (i.e., coherence) program~\cite{kang2017promising}, if $a$ is assigned $2$, then $b$ cannot be $1$, as that would violate SC-per-location since the write actions would be observed in different orders. 
Both examples can be automatically reasoned about with view-based protocols using VerCors-relaxed.\\[5pt]
\begin{figure}[t]
    \centering
	 {\begin{tabular}{c} 
			\begin{tabular}{l l || l l }
				\multicolumn{4}{c}{(2+2W)}                        \\
				$ x := 2 $ & & & $ y := 2 $ 				         \\
				$ y := 1 $ & & & $ x := 1 $ 						 \\
				$ a := y\ \textcolor{blue}{// 2} $ & & & $ b := x\ \textcolor{blue}{// 2} $ 						 \\
				\multicolumn{3}{c}{$T_0$}&\multicolumn{1}{c}{$T_1$}                              \\ 
			\end{tabular}$\ \ \ \ $
			
			\begin{tabular}{l l || l l }
				\multicolumn{4}{c}{(COH)}                                                 \\
				$ x := 1 $                         & & & $ x := 2$                        \\
				$ a := x\ \textcolor{blue}{// 2} $ & & & $ b := x\ \textcolor{red}{// 1}$ \\
				\multicolumn{3}{c}{$T_0$}&\multicolumn{1}{c}{$T_1$}                              \\ 
			\end{tabular}

	\end{tabular}}
    \caption{Two example programs using atomics with relaxed orderings.}
    \label{fig:intro_example}
\end{figure}

\noindent\textbf{Protocols.}
A protocol is a tree that models the stores to an atomic location for a particular thread. The concept comes from 
GPS~\cite{2014_GPS_turon}, in which all writes of a program to an atomic location are modeled in one protocol. The current 
work differs by defining a protocol per thread and atomic location combination. 
\begin{definition}
    A protocol $P^t_x$ is a tuple $(S_x^t, \hat{s}_x^t, \txrightarrow, F_x^t, V_x^t)$, where $x \in \ALoc$ with type $\mathbb{V}$ is an atomic location and $t \in \Tids$ a thread identifier, $S_x^t$ is a finite set of states, $\hat{s}_x^t \in S_x^t$ is the initial state, $F_x^t \subseteq S_x^t$ is the set of accepting states, $\rightarrow_{t,x} : S_x^t \times \Values \times S_x^t$ is the transition relation between states in $S_x^t$ with labels $v \in \Values$. The labels represent the values that are written to $x$, and doing so leads to the target state of the corresponding transition. Finally, $V_x^t : S_x^t \rightarrow \Values$ is a function that maps states to values, with $\forall (s, v, s') \in \rightarrow_{t,x}. V_x^t(s') = v$, and $V_x^t(\hat{s}_x^t) = v_0$, with $v_0$ the predefined initial value of $x$.
\end{definition}\vspace*{-2pt}
With $s \xrightarrow{v}_{t,x} s'$, we express that $(s, v, s') \in \rightarrow_{t,x}$. 
With $\rightarrow^*_{t,x}$, we refer to the transitive, reflexive closure of $\rightarrow_{t,x}$.
A protocol can be interpreted as a graph, with the states
being the vertices and $\rightarrow_{t,x}$ defining the edges. We restrict protocols to \emph{trees}, i.e., 1) $\forall s, s' \in S_x^t, v \in \Values. s 
\xrightarrow{v}_{t,x} s' \implies s' \not\rightarrow^*_{t,x} s$ ($\rightarrow_{t,x}$ is cycle-free) and
2) $\forall s, s' \in S_x^t. (s \not\rightarrow^*_{t,x} s' \wedge s' \not\rightarrow^*_{t,x} s) \implies \neg \exists s'' \in S_x^t. 
s \rightarrow^*_{t,x} s'' \wedge s' \rightarrow^*_{t,x} s''$ (for every two states between which no path defined by $\rightarrow^*_{t,x}$ exists, no state is reachable from both via $\rightarrow^*_{t,x}$).
Since protocols are trees, we can define a \emph{partial-order} of the states using $\rightarrow^*_{t,x}$:
for two states $s, s' \in S_x^t$,  we have $s \leq s'$ iff $s \rightarrow^*_{t,x} s'$. 



\begin{figure}[t]
    \centering
    \scalebox{0.75}{
        \includegraphics{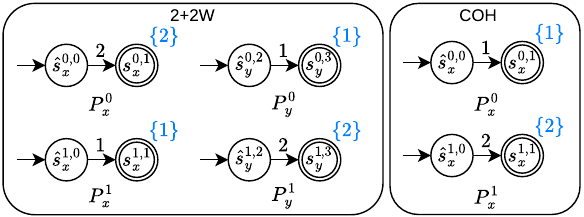}
    }
    \caption{Protocols for the 2+2W and COH examples in \cref{fig:intro_example}.}
    \label{fig:protocols}
\end{figure}

For the 2+2W program in \cref{fig:intro_example}, \cref{fig:protocols} shows four protocols, used to specify writing behavior. 
The protocols are used for speculative \getname operations, explained in the next section.
A protocol \smash{$P^t_x$} exists for each $t \in \Tids$ and $x \in \ALoc$ to which $t$ writes to. In the example, both threads write to $x$ and $y$, therefore we have \smash{$P_x^0, P_x^1, P_y^0\ \text{and}\ P_y^1$}.
 States are labeled \smash{$\stt{x}{i}{j}$}, where $i$ refers to thread $T_i$ associated with it, $j$ serves to differentiate states, and $x$ refers to the associated atomic location.
Initial states \smash{$\stti{x}{i}{j}$} are indicated by an incoming transition without a source state. Edges between states are part of 
 \smash{$\rightarrow_{t,x}$}, where the label signifies the written value. The values associated with the states are written in blue between curly brackets. The accepting states are indicated by double circles.

At the start of the parallel program, the protocols for all threads start in their initial states. At the end of the program, all protocols must have reached an accepting state for the execution to be valid w.r.t.\ the memory model. 

In the following, with $\States$, we refer to the set of all states of all protocols together for a given program, and with
$\States_x$, we refer to the set of all states of all protocols for location $x$. Similarly, with $\InitStates$, we refer to the set of
all initial states of all protocols, and with $\InitStates_x$ to the set of all initial states of protocols for $x$.

\vspace*{5pt}\noindent\textbf{Thread-local views.} 
\begin{definition}
Each thread $t$ has for every atomic location $x$ a \emph{thread-local view} 
\smash{$\Local_x^{t} : \Tids \rightarrow \States \cup \{ \bot \}$}.
The thread-local view maps each thread $t' \in \Tids$ to a \emph{state}
of its protocol $P_x^{t'}$. If the protocol does not exist, then $t'$ maps to $\bot$.
\end{definition}
Each thread $t$ uses for each location $x$ that it intends to access, the local view of its own protocol $P_x^t$, i.e., $\Local_x^t(t)$, to keep track of its own write
operations to $x$, and its local view of the protocols $P_x^{t'}$ of other threads $t'$, i.e., $\Local_x^t(t')$, to \emph{speculate} about the writes that those threads can
perform to $x$. 

Initially, each thread-local view points to the initial states, i.e., $\forall_{t, t' \in \Tids} \Local_x^{t}(t') = \hat{s}_x^{t'}$.
The set of all possible views for a program is called $\Views$.
For a given program with threads $\Tids$ and atomic locations $\ALoc$, the function $\Local: \Tids \times \ALoc \rightarrow \Views$ maps each pair $(t \in \Tids, x \in \ALoc)$ to thread-local view $\Local_x^t$.

The SOS rules in Fig.~\ref{fig:20:SOS} define how thread-local views evolve as \texttt{\getname} and \texttt{\setname} operations are executed. The contracts of \texttt{\getname} and \texttt{\setname} follow the SOS rules.
Prior to discussing the rules, we need to define which values a thread can read when a \texttt{\getname} operation is executed, as determined by a thread's local view. 

The values that can be read from location $x$ are the values that are associated with states 
from the protocols for $x$ that are reachable from the respective current states as defined by the view.
There are, however, two conditions: first, from its own protocols, a thread can only select the value associated with the current state,
not a value associated with a reachable state, i.e., it cannot speculate about its write operations, and second,
as soon as one write to $x$ has occurred, the initial value of $x$ cannot be read anymore.
In \cref{fig:20:PRead}, the functions \smash{$\PRead_x^t: \Views \rightarrow 2^\States$} map views to sets of reachable states, and
the functions $\Read_x^t: \Views \rightarrow 2^\States$ use this to involve the condition related to initial values.
Note the use of \Latest and \Time, which are there to keep track of the observed modification order of writes, to guarantee write and read coherence.
The most recently observed \texttt{\setname} operation, for every $x \in \ALoc$, represented by a protocol state, is stored by the function
$\Latest: \ALoc \rightarrow \States$, and the function $\Time: \ALoc \rightarrow \States \times \States$ maps locations to relations that store the observed modification order for that location. In $\PRead_x^t$, these are used to ensure that a state $s$ is selected only if it is not $\Time^+$-before (i.e. the transitive closure over $\Time$) the most recently observed \texttt{\setname} operation for that location. 

\begin{figure}[t]
    \centering
\[\PRead_x^t(\Local, \Latest, \Time) \equalhat
    \{s \mid \exists t' \in \Tids. \left( t \neq t' \wedge
    \Local_x^t(t') \rightarrow^*_{t',x} s \wedge \neg\Time(x)^+(s,\Latest(x)) \right)\} \cup \{ \pi(x) \}\! \]
    \[\Read_x^t(\Local, \Latest, \Time) \equalhat \left.
    \begin{cases}
        \PRead_x^t(\Local, \Latest, \Time), & \textrm{, if}\ \Latest(x) = \hat{s}_x^t  \\
        \PRead_x^t(\Local, \Latest, \Time) \setminus \{ \hat{s}_x^{t'} \mid \exists t' \in \Tids. \Local_x^{t}(t') \neq \bot \} & \textrm{, else}
    \end{cases}
    \right.\]
    \caption{The $\Read_x^t$ and $\PRead_x^t$ functions.}
\label{fig:20:PRead}
\end{figure}

In Fig.~\ref{fig:20:SOS}, the Write-rule expresses that a \setname operation can be executed when
from the current state $\Local_x^t(t)$ of $P_x^t$, a transition labeled $v$ exists to a state $s \in S_x^t$.
When this operation has been executed, the thread-local view is updated to indicate that the new current state
of $P_x^t$ is $s$. Also, $\Latest(x)$ is updated to $s$, and $\Time(x)$ is updated to indicate that the old $\Latest(x)$
is before $s$.
Program verification states are represented
by tuples $\langle \Local, \Latest, \Time \rangle$.

\begin{figure}[t]
    \centering
{\mbox{\inference[(Write)]
            {t \in \Tids \quad
                \Local_x^t(t) \xrightarrow{v}_{t,x} s \quad 
            }
            {\! \langle \Local, \Latest, \Time \rangle 
                \xrightarrow{x := v}_{t,x} 
                \langle \Local[(t,x) \mapsto \Local_x^t [t \mapsto s]], \Latest[x \mapsto s], \Time[x \mapsto \Time(x) \cup \{(\Latest(x), s)\}] \rangle\! }
        }
    \vspace{2ex}
    
        \mbox{\inference[(Read)]
            {    t, t' \in \Tids \quad
                s \in \Read_x^t(\Local, \Latest, \Time) \cap S_x^{t'} \quad
                x = V_x^{t'}(s)
            }
            {\!  \langle \Local, \Latest, \Time \rangle 
                \xrightarrow{a := x}_{t,x}  
                \langle \Local[(t,x) \mapsto \Local_x^t [ t' \mapsto s]], \Latest[x \mapsto s], \Time[x \mapsto \Time(x) \cup \{(\Latest(x), s)\}] \rangle\! \! 
            }
        }}
    \caption{Operational semantics for protocol views with \texttt{\getname} and \texttt{\setname} operations.}
    \label{fig:20:SOS}
\end{figure}


The Read-rule expresses that a \texttt{\getname} operation can be executed when a state $s \in S_x^{t'}$ can be read from according to
$\Read_x^t(\Local, \Latest, \Time)$.
We can then select $s$, which means that after execution of the \texttt{\getname} operation, the thread-local view and $\Latest(x)$ are updated,
and $\Time(x)$ is extended by mapping the old $\Latest(x)$ to $s$.

Consider the COH example in \cref{fig:intro_example}. The thread-local view $\Local_x^0$ is in its initial state with $\Latest(x) = \stti{x}{0}{0}$, $\Time=\{\}$ and \smash{$\PRead_x^0(\Local, \Latest, \Time) = \{\stti{x}{0}{0}, \stti{x}{1}{0}, \stt{x}{1}{1}\}$}, since both states of $P_x^1$ are reachable, i.e., \smash{$\forall s \in S_x^1. \Local_x^0(1) \rightarrow^*_{t',x} s$} holds and $\Time$ is empty.

After the write of 1 to $x$ by $T_0$, \smash{$\Local_x^0(0) = \stt{x}{0}{1}$} and \smash{$\Local_x^0(1) = \stti{x}{1}{0}$}. Suppose $T_0$ reads the value $2$ from \smash{$\stt{x}{1}{1}$} and $\Time$ is updated such that \smash{$(\stt{x}{0}{1}, \stt{x}{1}{1}) \in \Time$}. After the write of 2 to $x$ by $T_1$, \smash{$\Local_x^1(0) = \stti{x}{0}{0}$} and \smash{$\Local_x^1(1) = \stt{x}{1}{1}$}. In this state, \smash{$\PRead_x^1(\Local, \Latest, \Time) = \{\stti{x}{0}{0}, \stt{x}{1}{1}\}$}, since the both states of $P_x^0$ are reachable, however only for state $\stti{x}{0}{0}$ it holds that $\neg\Time(x)^+(s,\Latest(x))$, since $(\stt{x}{0}{1}, \stti{x}{1}{1}) \in \Time$.

The case distinction in $\Read_x^t$ is to differentiate a local view from which the initial state is observable. If some thread $t$ has observed any non-initial state, i.e., $\exists t'\in\Tids. \Local_x^t(t') \neq \hat{s}_x^t$, it means that the initial state can never be observed anymore, since for any non-initial state $s$ it holds that $\hat{s}_x^t \leq s$. This corresponds to the second case in $\Read_x^t$, where the initial states are removed if $\Latest(x) \neq \hat{s}_x^t$. Do note here that all initial states essentially represent the same starting value for an atomic location, e.g., $\stti{x}{0}{0}$ and $\stti{x}{1}{0}$ in \cref{fig:protocols} both represent the initial state of $x$. Although they represent the same starting value, they are differentiated since they are associated with different threads.

%
%
%
%
%
%
%

\vspace*{5pt}\noindent\textbf{The Verification Procedure.}
This section explains the verification procedure of parallel programs using the view-based protocols.
The encoding of the procedure into VerCors-relaxed is explained in \cref{sec:40:encoding} and \cref{sec:gencontract}\ifextended. \fi
At the start, for all threads $t \in \Tids$, an initial thread-local view $\Local_x^t$ is created for each $x \in \ALoc$,
the function $\Latest$ is initialized to the respective initial states for each atomic location and the function $\Time$ is created, initially undefined.
Then, repeatedly, the SOS rules are applied in all possible ways, to obtain a state space of executions.
For the construction of an execution, first,
the operations of one thread $t$ are considered in program order, having first defined $\Latest(x) = \hat{s}_x^t$ for all $x \in \ALoc$, or
$\Latest(x) = \bot$, if the protocol does not exist.
The $\Local_x^t$ for all relevant $x \in \ALoc$, $\Latest(x)$ and $\Time(x)$ are updated according to the SOS rules as the
corresponding operations are executed.
Once this has been done, the operations of another thread $t' \neq t$ are considered,
after $\Latest$ has been reset, this time reading and updating the
$\Local_x^{t'}$ views. When all threads have been processed in this way, a final consistency check is performed to determine
whether an execution speculated a value that was never written to: if this is the case, the execution is ignored.
An execution is consistent iff:
\begin{equation}
    \forall t \in \Tids, x \in \ALoc. \Local_x^t(t) \neq \bot \implies \Local_x^t(t) \in F_x^t \tag{CC1}
\end{equation}
\begin{equation}
    \forall t, t' \in \Tids, x \in \ALoc. \Local_x^t(t') \rightarrow^* \Local_x^{t'}(t') \tag{CC2}
\end{equation}

\noindent Condition CC1 expresses that in each protocol, we have reached a final state in the view of the thread owning that protocol.
A violation of this would mean that the thread has not completely behaved as specified by the protocol. The condition CC1 has to hold at the end of the program. The condition CC2 expresses that
any speculations done by a thread $t$ about writes of a thread $t'$ have been valid, i.e., the state $\Local_x^{t'}(t')$, representing
the actual writes of $t'$, is reachable from $\Local_x^t(t')$, representing the writes that $t$ assumed $t'$ would
perform. The condition CC2 has to hold throughout the execution.

To demonstrate this procedure, we consider again the examples in \cref{fig:intro_example}.
    Consider the 2+2W example, for which the protocols given in \cref{fig:protocols}.
    Initially, we have \smash{$\Local_x^0(0) = \stti{x}{0}{0}$}, \smash{$\Local_x^0(1) = \stti{x}{1}{0}$},
    \smash{$\Local_y^0(0) = \stti{y}{0}{2}$} and \smash{$\Local_y^0(1) = \stti{y}{1}{2}$}. Views \smash{$\Local_x^1$} and $\Local_y^1$ are
    defined similarly.
    When verifying $T_0$'s instructions, we first have $\Latest(x) = \stti{x}{0}{0}$ and \smash{$\Latest(y) = \stti{y}{0}{2}$},
    and then encounter "\smash{$x:=2$}", resulting in \smash{$\Local_x^0(0) = \stt{x}{0}{1}$}, $\Latest(x) = \stt{x}{0}{1}$ and
    \smash{$\Time(x) = \{ (\stti{x}{0}{0}, \stt{x}{0}{1}) \}$}.
    Next, "$y := 1$" is encountered, resulting in \smash{$\Local_y^0(0) = \stt{y}{0}{3}$, $\Latest(y) = \stt{y}{0}{3}$}, and
    \smash{$\Time(y) = \{ (\stti{y}{0}{2}, \stt{y}{0}{3}) \}$}.
    After that, a \texttt{\getname} operation from $y$ is encountered. The possible values are represented by $\Local_y^0(0)$, plus all states
    reachable from \smash{$\Local_y^0(1)$}, minus the latter's initial state, as a write to $y$ has been observed (see the
    definition of $\Read_x^t$). This means that the possible values are \smash{$V_y^0(\stt{y}{0}{3}) = 1$} and \smash{$V_y^1(\stt{y}{1}{3}) = 2$}.
    Hence, operations "$y := 1$" and "$y := 2$" by $T_0$ are both enabled. We set $\Latest(y) = \stt{y}{1}{3}$ and
    \smash{$\Time(y) = \{ (\stti{y}{0}{2}, \stt{y}{0}{3}), (\stt{y}{0}{3}, \stt{y}{1}{3}) \}$}. Next, we consider $T_1$. The first two \texttt{\setname} operations lead to \smash{$\Local_y^1(1) = \stt{y}{1}{3}$}
    and \smash{$\Local_x^1(1) = \stt{x}{1}{1}$} (and appropriate updates of $\Latest$ and $\Time$).
    In a similar way as for $T_0$, the next two possible \texttt{\getname} operations are associated with
    $\stt{x}{0}{1}$ and $\stt{x}{1}{1}$, meaning that both $1$ and $2$ can be read. Hence, the result indicated in \cref{fig:intro_example}
    is consistent.

    Consider the COH example in \cref{fig:intro_example}, and its protocols in \cref{fig:protocols}.
    Initially, we have $\Local_x^0(0) = \Local_x^1(0) = \stti{x}{0}{0}$ and $\Local_x^0(1) = \Local_x^1(1) = \stti{x}{1}{0}$.
    After "$x := 1$" by $T_0$, $\Local_x^0(0) = \stt{x}{0}{1}$, $\Latest(x) = \stt{x}{0}{1}$, and $\Time(x) = \{ (\stti{x}{0}{0}, \stt{x}{0}{1}) \}$.
    Next, according to $\Read_x^t$, a \texttt{\getname} operation can access
    the results represented by both $\stt{x}{0}{1}$ and $\stt{x}{1}{1}$. We select $\stt{x}{1}{1}$, to check whether the result indicated in \cref{fig:intro_example}
    is consistent, and set $\Latest(x) = \stti{x}{1}{1}$ and $\Time(x) = \{ (\stti{x}{0}{0}, \stt{x}{0}{1}), (\stt{x}{0}{1}, \stt{x}{1}{1}) \}$. Next,
    after $\Latest(x)$ has been reset to $\stti{x}{1}{0}$, "$x := 2$" by $T_1$ leads to
    $\Local_x^1(1) = \stt{x}{1}{1}$, $\Latest(x) = \stti{x}{1}{1}$, and $\Time(x) = \{ (\stti{x}{0}{0}, \stt{x}{0}{1}), (\stt{x}{0}{1}, \stt{x}{1}{1}), (\stti{x}{1}{0}, \stt{x}{1}{1}) \}$.
    As
    $\Time(x)^+(\stt{x}{0}{1}, \stt{x}{1}{1})$, the next \texttt{\getname} operation can only read from $\stt{x}{1}{1}$, meaning that only "$x := 2$" by $T_1$ is enabled, and hence
    the result in \cref{fig:intro_example} is inconsistent since it violates read coherence.

\section{Relating View-based Protocols to SLR}
\label{sec:encoding_slr}

SLR is a program logic for reasoning about concurrent programs under weak memory models. Its key feature is its reasoning about relaxed memory accesses by tracking the reads and writes of the program and the order of observation of values. Using this information, invalid executions can be shown to be impossible.

We relate our view-based protocols to the relaxed fragment of the SLR logic rules, namely the $\wrlx$, $\rrlx$ and $\rrlxs$ rules, which are introduced below\footnote{Details pertaining to other memory orderings such as $Rel$ and $Acq$ assertions and modalities are omitted to focus on the relaxed memory ordering.}.

Besides the standard operators in first order logic and separation logic, the assertion language of SLR introduces two novel assertions, namely $O(l,v,ts)$ and $W^p(l,X)$. The assertion $O(l,v,ts)$ signifies the observation of value $v$ from location $l$ at time $ts$. The timestamp $ts$ is used to order observations where the timestamps are densely totally ordered. 

The assertion $W^p(l,X)$ signifies permission to write to a location $l$, where $X$ is a set of pairs $(v,ts)$ where a value $v$ is written at timestamp $ts$. The fractional permission $p$ signifies whether the permission is exclusive (i.e., $p=1$) or shared (i.e., 0 < $p$ < 1). This assertion is encoded using the protocols and views. In the proposed approach, the $p$ is not explicitly encoded, but encoded implicitly by the protocols. The case where there is only one thread writing, meaning there is only one protocol with an edge, implies exclusive access, i.e., $p = 1$. In the case where multiple threads write, meaning multiple protocols in a local view can progress, implies shared accesses, i.e., $0 < p < 1$. 

\begin{figure}[t]
    \vspace*{-6pt}
    {\centering\(\displaystyle
        \hfill
        \frac{\phi(v)\ \text{is pure}}
        {\;\vdash
            \left\{
            \begin{array}{l}
                W^p(x, X)\ * \phi(v)\ * \\
                \;\;\;  O(x,-,ts)
            \end{array}
            \right\}
            \ x := v
            \left\{
            \begin{array}{l}
                \exists ts' > ts. \\
                W^p\bigl(x,\{(v,ts')\}\cup X\bigr)
            \end{array}
            \right\}
        }
        \quad\text{($\wrlx$)}
        \)
    }
    \caption{The $\wrlx$ rule}
    \label{fig:25:wrlx}
\end{figure}

\vspace*{5pt}\noindent\textbf{The $\mathbf{\wrlx}$ rule.} 
The $\wrlx$ rule in \cref{fig:25:wrlx} states that the threads need to have permission to write to $x$ (i.e., $W^p(x, X)$) and a previously observed value at time $ts$ (i.e.,  $O(x,-,ts)$) where the observed value is irrelevant. Also, $\phi(v)$ has to hold, where $\phi$ is a pure predicate over $v$. It has to be pure, since relaxed memory operations cannot by themselves be used for synchronization\footnote{In SLR's release/acquire fragment, $\phi$ is used for resource sharing, since release/acquire semantics can be used for synchronization. This work focuses on SLR's relaxed fragment, so the encoding of $\phi$ is not shown to focus on the $W$ and $O$ assertions.}. This predicate can then be communicated to a thread that observes the written value. After the write operation, the set of writes $X$ is updated with the newly written pair $(v, ts')$ where the timestamp $ts'$ is chosen such that it is strictly larger than the timestamp $ts$ of the previously observed value.

SLR uses timestamps in its $W$ and $O$ assertions. The timestamps are used to record the order of writes and observations, where the concrete timestamps are irrelevant. For this reason, the view-based protocols do not encode the timestamps explicitly, but model the modification order using \Time. This relation defines an order between \texttt{\setname} operations where there is a concrete order. For example, the \Time relation is used in the $\PRead_x$ function to only read values that are not \Time-before the most recently observed \texttt{\setname} operation, i.e., a value is observed that is not followed by a \texttt{\setname} operation with a larger timestamp. 
This is also used to order the \texttt{\setname} operations per-thread, using the protocols to define the program order per-thread, where every edge $(s \rightarrow^*_{t,x} s')$ defines $(s,s') \in \Time(x)^+$.

The SOS Write rule (in \cref{fig:20:SOS}) explains how the local views change when a value is written to an atomic location. To explain the link with the $\wrlx$ rule, we must first discuss the encoding of the assertions $W^p(x,X)$ and $O(x,v,ts)$. The set $X$ in the $W^p(x,X)$ assertion is conceptually a history of the observed \texttt{\setname} operations for location $x$ which are ordered by timestamps. This is in essence the same concept as $\Time(x)$, however encoded differently. 

The set $X$ is updated in the postcondition of the $\wrlx$ rule by adding $(v',ts')$ to $X$, where $v'$ is the value to write and $ts'$ is a timestamp that is later than the most recent observation $O(x,-,ts)$. This corresponds to the update of local views in the Write-rule, where for the thread $t$ and a reachable state $s$ (with $V_x^{t}(s) = v$), the local view is updated, i.e., $\Local_x^{t} [{t} \mapsto s]$. Also, $\Latest$ is updated, i.e., $\Latest[x \mapsto s]$, where $\Latest(x)$ corresponds to the \texttt{\setname} operation associated with $O(x,v,ts)$. The order between the state of the most recently observed \setname $\Latest(x)$ and the state of the current \texttt{\setname} operation is added to $\Time(x)$. This encodes the coherence property Writes-Ordered in \cref{fig:25:writes-ordered}, where the order beforehand was left undefined and after the \texttt{\setname} operation, the order is established.

\begin{figure}[t]
{
\begin{flalign*}
W&^p(x,X) * (-,ts) {\in} X * (-,ts') {\in} X \Rightarrow \\
 &W^p(x,X)\ * (ts < ts' \lor ts = ts' \lor ts' < ts) \tag{Writes-Ordered}\\[5pt]
W&^{1}(x,X)\ *\ O(x,v,ts)\ \;\Rightarrow\;\\
 &W^{1}(x,X)\ *\ O(x,v,ts)\ *\ (v,ts') \in X \tag{Reads-from-Write}\\[5pt]
W&^p(x,X)\ *\ (v,ts) \in X\ *\ (v',ts') \in X\ *\ v \neq v'\ \;\Rightarrow\;\\
 &W^p(x,X)\ *\ ts \neq ts' \tag {Different-Writes}\\
\end{flalign*}  
}\vspace*{-30pt}
    \caption{The coherence properties}
    \label{fig:25:writes-ordered}
\end{figure}

The $O(x,v,ts)$ assertion is encoded mainly using $\Latest$, where the most recently observed \texttt{\setname} operation is recorded. The order between observations $O(x,v,ts)$ and $O(x,v',ts')$ is encoded using $\Time(x)$, where for two states $s$ and $s'$ with $V_x(s) = v$ and $V_x(s')= v'$, if $ts<ts'$, then $(s,s') \in \Time(x)$.

\begin{figure}[t]
\vspace*{-12pt}
\centering
\begin{flalign*}
\vdash &\{\, O(x,-,ts) \,\}\ r := x\ \{\, \exists ts' \ge ts.\ O(x,v,ts') * r = v * \phi(v) \,\} \tag{\rrlx} \\[5pt]
\vdash &\{\, W^1(x,X) \,\}\\
       &\ \ r := x\\
       & \{\, \exists ts.\ (v,ts) = \max(X) * W^1(x,X)\ * r=v * O(x,v,ts) * \phi(v) \}\tag{\rrlxs}
\end{flalign*}
\caption{The $\rrlx$ and $\rrlxs$ rules}
\label{fig:25:rlx}
\end{figure}

\noindent\textbf{The $\mathbf{\rrlx}$ and $\mathbf{\rrlxs}$ rules.}
The $\rrlx$ rule in \cref{fig:25:rlx} states that a thread again needs a previously observed value at time $ts$ (i.e.,  $O(x,-,ts)$). After reading from $x$, the value $v$ is observed at a timestamp $ts'>ts$, and the predicate $\phi(v)$ can be established (dually from the $\wrlx$ rule). 

In the Read-rule (in \cref{fig:20:SOS}), the state $s$ is observed and the local view consequently updated, i.e., $\Local_x^t [ t' \mapsto s]$, along with $\Latest$ and $\Time(x)$. Remember that the local view represents the most recently observed \texttt{\setname} operation of each thread, which implies that if state $s$ is the result of the \texttt{\getname} operation, then the local view has observed a thread writing the value associated with $s$. This encodes the Reads-from-Write property in \cref{fig:25:writes-ordered}, since the result of the \texttt{\getname} operation must have the associated \texttt{\setname} operation in the local view. The property Reads-from-Write states that an observation (i.e. a read) of value $v$ at time $ts$ implies that there is a write before where $v$ has been written to $x$ at time $ts'<ts$.

%

When a thread has exclusive write permission, more conclusions can be drawn from an observation, expressed in the $\rrlxs$ rule. Before the \getname operation, exclusive write permission is required (i.e., $W^1(x,X)$). Afterwards, again, the observation of value $v$ is made at time $ts$ and the predicate $\phi(v)$ can be established. Moreover, since the write permission is exclusive, it is established that the value $v$ with timestamp $ts$ is the maximum value of $X$, i.e., the latest write.

The encoding of this rule is subtle. 
The Read rule returns a value associated with state $s$, where $s \in \Read_x^t(\Local, \Latest, \Time) \cap S_x^{t'}$. 
The $\PRead$ function (which is called by the $Read$ rule) returns the set of possibly observable states where the state $s$ is in the set iff $\Local_x^t(t') \rightarrow^*_{t',x} s$, i.e., it is reachable from the current thread-local view. 
If there is only one writing thread $t$, i.e., one protocol $P_x^t$, the conclusion can be drawn that the only reachable state is in $P_x^t$. If thread $t$ then performs a \texttt{\getname} operation, it can conclude that the result stems from the most recent \texttt{\setname} of itself, which is equivalent to the maximum of the set $X$. If there are multiple writers, this cannot be deduced as with the $\rrlx$ rule.

\vspace*{5pt}\noindent\textbf{Coherence.}
SLR allows to reason about coherence with the rules Reads-from-Write, Writes-Ordered and Different-Writes. The first two rules were mentioned above. The property Different-Writes in \cref{fig:25:writes-ordered} is captured by the protocols. The protocol states represent \texttt{\setname} operations by different threads, even if two threads write the same value. SLR defines this using the written values, where given the assertion $W^p(x,X)$, two \texttt{\setname} operations in the set $X$ with different values happen at different times. The protocol states make this explicit by associating distinct states with both \texttt{\setname} operations. This property is maintained by the Write-rule by tracking the latest observed \texttt{\setname} operation $\Latest(x)$ and establishing the order of the current \texttt{\setname} operation and $\Latest(x)$. 

\vspace*{5pt}\noindent\textbf{Soundness.}
Soundness in this context of this work means that if a program encoded using the view-based protocols verifies, then the program also verifies with the SLR logic. 
The argument is made as follows:

\begin{compactenum}
    \item For the program verification state $\langle \Local, \Latest, \Time \rangle$, it initially holds that $\Latest(x) = \hat{s}_x$, \smash{$\forall_{t, t' \in \Tids} \Local_x^{t}(t') \equalhat \hat{s}_x^{t'}$}, and $\Time$ is undefined. This corresponds with SLR's initial $W(x, X)$ where $X = \{(\hat{s}_x,0)\}$. 
    
    \item For a thread $t$ and location $x$, the SOS Write rule updates the program state by 1) $\Local[(t,x) \mapsto \Local_x^t [t \mapsto s]]$, updating the view to point to the next state $s$ that corresponds with the \setname operation and 2)  $\Time[x \mapsto \Time(x) \cup \{(\Latest(x), s)\}]$, establishing the order between the most recently observed write operation and $s$.
    This corresponds to the addition of the pair $(v, ts')$ to the set $X$ in SLR's $W$ assertion, i.e, $W(x, \{(v,ts')\} \cup X)$ where $V^t_x(s)=v$ and $ts'>ts$. 
    
    \item For a thread $t$ and location $x$, the SOS Read rule updates $\Local^t_x[t' \mapsto s]$, where $s$ is a state representing a possibly observable write operation (of thread $t'$). The order of the previously observed operation $\Latest(x)$ and $s$ is added to the $\Time$ relation. This corresponds to the $O(x,v,ts')$ assertion in SLR's $\rrlx$ and $\rrlxs$ rules, where $ts'>ts$ corresponding to the update to $\Time$. 
    
    The case where there is only one writing thread, i.e., one protocol, the set  $\Read_x^t(\Local, \Latest, \Time) \cap S_x^{t'}$ from which $s$ is chosen, is limited to the states of the writing protocol. This corresponds to the additional postcondition in the $\rrlxs$ rule where $(v,ts) = max(X)$.
    
    \item From an initial program state $\langle \Local, \Latest, \Time \rangle$ and corresponding $W(x,X)$ and $O(x,-,ts)$ assertions, we can repeated apply the SOS rules and SLR rules to end up in a final program state $\langle \Local', \Latest', \Time' \rangle$. Throughout the application of the rules, we can maintain a correspondence between $\langle \Local, \Latest, \Time \rangle$ and the $W(x,X)$ and $O(x,-,ts)$ assertions, as explained above.
    From this we conclude that if the final program state $\langle \Local', \Latest', \Time' \rangle$ is reached, the $W(x,X')$ and $O(x,-,ts')$ assertions are updated accordingly, implying that if a program verifies with the view-based protocols approach, then also the program verifies with the SLR logic. Thus, we conclude that our approach is sounds w.r.t. SLR.
\end{compactenum}
%
%

\section{Encoding Protocols and Views into VerCors-relaxed}
\label{sec:40:encoding}


In this section, we explain parts of the encoding of the protocols and thread-local views into VerCors-relaxed. 
For presentation purposes, we omit the contract template for parallel programs (see~\cref{sec:gencontract}). Furthermore, we leave out the encoding for $\Time$
in the method contracts for \getname and \setname, addressing the SOS Read and Write rules, respectively. 

Protocols are encoded using VerCors' ADTs (axiomatic data types). The states and transitions of the protocols are defined axiomatically because VerCors' ADT support allows the modeling of abstract concepts, similar to how (concrete) lists are modeled as an abstract ordered collection. The ADT consists of three elements: a type name, abstract functions and axioms. 

\begin{figure}[t]
\begin{lstlisting}[basicstyle=\ttfamily\footnotesize]
adt ps {#\label{lst:30:adt_prot:adt}#
	pure ps sx0(int t);#\label{lst:30:adt_prot:st0}# // State definitions
	pure ps sx1(int t);
	... // #\frl# states #\label{lst:30:adt_prot:st1}#	
	pure #$\Values$# val(ps s);#\label{lst:30:adt_prot:val}#
	axiom val(sx0(0)) == a && val(sx1(0)) == b && ...; #\label{lst:30:adt_prot:val_axioms}# // #\frl# states with concrete a and b #\label{lst:30:adt_prot:val_axioms_1}#
	pure boolean isF(int t, ps s);#\label{lst:30:adt_prot:accept}#
	axiom isF(0, s) == (s == sx1(0)) && .. ; // #\frl# threads#\label{lst:30:adt_prot:accept_axioms}#
	pure ps next(#$\Values$# val, ps s);#\label{lst:30:adt_prot:next}#
	pure int intra(ps s);#\label{lst:30:adt_prot:intra}#
	axiom next(b, sx0(0)) == sx1(0) && ...; // #\frl# transitions#\label{lst:30:adt_prot:next_ex}#
	axiom (#\frl# int i, ps s; intra(s) < intra(next(i, s)));#\label{lst:30:adt_prot:intra_next_axiom}# }
	\end{lstlisting}
	\caption{Skeleton code for a protocol encoding.}
	\label{fig:30:adt_protocol}
\end{figure}

\cref{fig:30:adt_protocol} shows skeleton code of the ADT encoding named \texttt{ps} (for protocol states, \cref{lst:30:adt_prot:adt}). This introduces the \texttt{ps} type. For a given protocol $P_x^t$, and for each state in $S_x^t$, a unique abstract function is defined (\crefrange{lst:30:adt_prot:st0}{lst:30:adt_prot:st1}) that acts as the constructor for that state, with a thread identifier the state belongs to.
$V_x^t$ is encoded as a function \texttt{val} (\cref{lst:30:adt_prot:val}). For each state, the associated value is defined (\cref{lst:30:adt_prot:val_axioms}). The accepting states $F_x^t$ are encoded as a function \texttt{isF}, taking a thread identifier and a state (\cref{lst:30:adt_prot:accept}).
At (\cref{lst:30:adt_prot:accept_axioms}), the accepting states in $F_x^t$ are defined, e.g., \texttt{sx1(0)} is accepting for thread \texttt{t}.
For all states in $F_x^t$, the result of \texttt{isF} is defined. For the other states, the result is undefined,
resulting in a failing verification.
 
Finally, $\txrightarrow$ is encoded using two functions. First, the next function (\cref{lst:30:adt_prot:next}) takes a state and a value and returns the next state. Each transition is defined at \cref{lst:30:adt_prot:next_ex}, e.g., the next state of \texttt{sx0(0)} is \texttt{sx1(0)}
 when writing the value \texttt{b}. 

The \texttt{intra} function (\cref{lst:30:adt_prot:intra}) defines a mapping of states to integers, used to define the partial order between states. For each state $s$, the intra-value of any next state of $s$ is greater than the intra-value of $s$. Note that since the \texttt{next} function is not total, the result of \texttt{intra} is also not total.

\cref{fig:30:contract_cons,fig:30:contract_get,fig:30:contracts_set} shows contracts for the atomic methods.
Here, a global view \texttt{G} is used, which is an array of \texttt{ps}, where $\forall {t\in\Tids} . \texttt{L[t][t]} = \texttt{G[t]}$.
\texttt{G} is used as a shorthand to simplify notation.
The atomic location is represented by a class \texttt{a\_V} (\cref{fig:30:contract_cons}, \cref{lst:30:a_V}).
The constructor returns the permissions over the global and local views (\cref{lst:30:a_V:perm_pss,lst:30:a_V:perm_vs}) and initializes the global and local views (\crefrange{lst:30:a_V:init_vs}{lst:30:a_V:init_pss}).

\begin{figure}[t]
\begin{lstlisting}[basicstyle=\ttfamily\footnotesize]
class a_V {#\label{lst:30:a_V}#
	ps[] G;
	seq<ps>[] L;
	/*@ ensures Perm(G[*], write) ** Perm(L[*], write);#\label{lst:30:a_V:perm_pss}##\label{lst:30:a_V:perm_vs}#
		   ensures (#\frl# int t1,t2=0..N; L[t1][t2] == s0(t2));#\label{lst:30:a_V:init_vs}#
		   ensures (#\frl#* int t1=0..N; G[t1] == s0(t1));#\label{lst:30:a_V:init_pss}# @*/
	constructor(int N); }
	\end{lstlisting}
	\caption{The contract of the constructor.}
	\label{fig:30:contract_cons}
\end{figure}

The contract of the \getname method (\cref{fig:30:contract_get}) defines the relations between the local view before the method call, the view after the method call and the global view. To express these, \getname takes four ghost arguments (i.e., the \texttt{given} arguments) and returns two ghost results (i.e., the \texttt{yields} results). These are specification-only variables that are not part of the program, but help to reason
about the problem. The pre-/postconditions regarding protocol adherence of the views are omitted due to space limitations. The \texttt{t} and \texttt{N} specify the thread accessing the atomic location and the number of threads, respectively. The \texttt{oldLV} and \texttt{newLV} are the local view before and after the \getname method. The \texttt{prevPi} and \texttt{nextPi} ghost variables are discussed later.

\begin{figure}[t]
\begin{lstlisting}[basicstyle=\ttfamily\footnotesize]
/*@ given seq<ps> oldLV;#\label{lst:30:a_V:gst_start}#
	 yields seq<ps> newLV;
	 given int N, t; 
	 given ps prevPi;
	 yields ps nextPi; #\label{lst:30:a_V:gst_witt}#
 #\textcolor{red}{1)}#context Perm(G[*], 1\2); #\label{lst:30:a_V:pss_read_perm}#
	 ensures (#\frl#* int t1=0..N; G[t1] == \old(G[t1])); #\label{lst:30:a_V:pss_no_change}#
 #\textcolor{red}{2)}#ensures newLV[t] == \old(oldLV[t]);  #\label{lst:30:a_V:globlocsame_2}#
	 requires (#\frl# int t1,t2=0..N; intra(oldLV[t1][t2]) <= intra(G[t2])#\label{lst:30:a_V:ginvoldLV}#
															|| intra(G[t2]) <= intra(oldLV[t1][t2])));#\label{lst:30:a_V:ginvoldLV2}#
	 ensures(#\frl# int t1,t2=0..N; intra(newLV[t1][t2]) <= intra(G[t2]) #\label{lst:30:a_V:ginvnewLV}#
														||  intra(G[t2]) <= intra(newLV[t1][t2])));#\label{lst:30:a_V:ginvnewLV2}#
	 ensures (#\frl# int t1=0..N; intra(oldLV[t1]) <= intra(newLV[t1])); #\label{lst:30:a_V:updateNewLV}#
 #\textcolor{red}{3)}#ensures \result \in Read(newLV) && ps.val(nextPi) == \result; #\label{lst:30:a_V:get_next_val}##\label{lst:30:a_V:get_next_res}#
	 ensures (prevPi != nextPi) ==> ps.tau(prevPi, nextPi);@*/ #\label{lst:30:a_V:load:prev_next_pi}#
int load();
	\end{lstlisting}
	\caption{The contract of the \getname method.}
	\label{fig:30:contract_get}
\end{figure}

The contract of the \getname method is split into three parts:

\begin{compactenum}
	\item \textbf{Permission annotations}: To read from the protocol of a different thread, \getname requires thread read permission to all global views (\cref{lst:30:a_V:pss_read_perm}). A thread acquires this permission from a global invariant (at the call site of \getname). It is specified that the global view does not change (\cref{lst:30:a_V:pss_no_change}), since the \getname method only requires read permission and does not change the global view.	
	
	\item \textbf{Relation between global view and the local views}: 
	First, \cref{lst:30:a_V:globlocsame_2} specifies that the old and new local views of the accessing thread are consistent. 
	Next, \crefrange{lst:30:a_V:ginvoldLV}{lst:30:a_V:ginvnewLV2} maintain the invariant between global view and \texttt{oldLV}/\texttt{newLV}, respectively.
	Finally, \cref{lst:30:a_V:updateNewLV} specifies how \texttt{newLV} can change w.r.t.\ \texttt{oldLV}, where each state in \texttt{newLV} has to be reachable from the respective state in \texttt{oldLV}.
	 
	\item \textbf{Relation between local views and the result}: The \texttt{Read} function (\cref{lst:30:a_V:get_next_val}) corresponds to $\Read_x^t$ from \cref{sec:20:protocols}. The set of possible values for \texttt{newLV} must contain the returned result corresponding to the value of most recently observed state \texttt{nextPi}. The order between the most recently observed state \texttt{nextPi} and the previously observed state \texttt{prevPi} is also established (\cref{lst:30:a_V:load:prev_next_pi}).
\end{compactenum}
 
\begin{figure}[t]
\begin{lstlisting}[basicstyle=\ttfamily\footnotesize]
/*@ given int N, t;
	 given seq<ps> oldLV; 
	 yields seq<ps> newLV; 
	 given ps prevPi;
	 yields ps nextPi;
 #\textcolor{red}{1)}#context Perm(G[t], 1\2) ** Perm(G[t], 1\2); #\label{lst:30:a_V:set_perms}#
 #\textcolor{red}{2)}#requires ps.val(ps.next(val, oldLV[t])) == val;#\label{lst:30:a_V:set_next_is_val}#
 #\textcolor{red}{3)}#ensures G[t] == ps.next(val, oldLV[t]);#\label{lst:30:a_V:set_up1}#
	 ensures newLV == oldLV[t -> ps.next(val, oldLV[t])];#\label{lst:30:a_V:set_up2}#
	 ensures ps.intra(newLV[t]) <= ps.intra(G[t]);#\label{lst:30:a_V:set_up3}#
 #\textcolor{red}{4)}#ensures \result == val ** ps.val(nextPi) == val; 
	 ensures ps.tau(prevPi, nextPi); @*/#\label{lst:30:a_V:store:prev_next_pi}#
int store(int val);
\end{lstlisting}
	\caption{The contract of the \setname method.}
	\label{fig:30:contracts_set}
\end{figure}

The contract for the \setname method (\cref{fig:30:contracts_set}) has similar ghost arguments/results as \getname, where the \getname method mainly changes its local view of the progress of other threads, the \setname method only records its own progress. The contract consists of four parts:

\begin{compactenum}
	\item \textbf{Permission annotations}: Only write permission is needed to \texttt{t}'s global view. Half of the permission comes from a global invariant (at the call site of \setname) and the other half from the thread itself (\cref{lst:30:a_V:set_perms}).
	
	\item \textbf{Valid transition}: The argument \texttt{val} must be the value of a next state of the protocol (\cref{lst:30:a_V:set_next_is_val}). 
	
	\item \textbf{Update local view and global view}: The global view for \texttt{t} is set to the next state (\cref{lst:30:a_V:set_up1}), and \texttt{newLV} is copied from \texttt{oldLV}, with \texttt{newLV[t]} changed accordingly (\cref{lst:30:a_V:set_up2}). Finally, it is maintained that the global view and new local view uphold the partial order without speculations (\cref{lst:30:a_V:set_up3}).
	
	\item \textbf{Relation between local state and result}: The relation between the local state and result is similar to the same relation in the \getname method. 
\end{compactenum}

\iftrue
\noindent A contract of the \casname operation can be proven by combining the contracts for \getname and \setname. Due to space limitations, the section on \casname operations has been moved to \cref{app:cas}\ifextended.\fi

\else
Finally, \cref{fig:30:contracts_cas} shows the contract for the \casname method. The contract of the \casname operation is join the contracts for \getname and \setname. With the contracts for the abstract \getname and \setname methods, the implementation of the \casname method can be verified. For brevity, overlapping with the contraints of \getname and \setname are omitted to focus on the differences. The contract mainly consists of two parts:

\begin{compactenum}
	\item \textbf{\casname fail}: When the operation fails, the global view of \texttt{t} does not change (\cref{lst:30:cas:nochange}). The call to \getname updates the local view. The return value of \getname cannot be equivalent to \texttt{oldVal} (\cref{lst:30:cas:newval_no_oldval}).
	Finally, a description of \texttt{newLV} can be given (\cref{lst:30:cas:descriptionfail}). The predicate \texttt{descriptionFail} is a placeholder and is to be defined by the user to use in further proofs. It can be used to exclude certain executions, see \cref{casexample}.
	
	\item \textbf{\casname success}: The \texttt{newLV} is specified to have changed similarly to the \setname method's contract. In addition to this, a token \texttt{CASSuccess} is given to signify that the operation has succeeded (\cref{lst:30:cas:cassucces}). This token captures the state that \texttt{oldVal} is read from. The token is used to exclude multiple operations reading from the same write.
	Again, a description of \texttt{newLV}  can be given (\cref{lst:30:cas:descriptionsucceed}) with the predicate \texttt{descriptionSuccess}.
\end{compactenum}

\begin{figure}[t]
	\begin{lstlisting}		
/*@ given seq<ps> oldLV; 
  yields seq<ps> newLV;
  given int N, tid, depTid, prev;
  yields int next; 
#\textcolor{red}{1)}#ensures !\result ==> G[tid] == \old(G[tid]);#\label{lst:30:cas:nochange}#
  ensures !\result ==> next != oldVal; #\label{lst:30:cas:newval_no_oldval}#
  ensures !\result ==> descriptionFail(newLV);#\label{lst:30:cas:descriptionfail}#
#\textcolor{red}{2)}#ensures \result == G[tid] == ps.next(oldLV[tid]);
  ensures \result ==> newLV[tid] == ps.next(oldLV[tid]);
  ensures \result ==> ps.val(newLV[tid]) == newVal;
  ensures \result ==> next == newVal;
  ensures \result ==> next \in Read(newLV);
  ensures \result ==> CASSuccess(tid,newLV[depTid]);#\label{lst:30:cas:cassucces}#
  ensures \result ==> descriptionSuccess(newLV);#\label{lst:30:cas:descriptionsucceed}#@*/
boolean rmw(int oldVal, int newVal) {#\label{lst:30:a_V:cas}#
	int c = get() given {...} yields {...};
	if (c == oldVal) { return true; } 
	return false; 
}
	\end{lstlisting}
	\caption{The contract of \casname method.}
	\label{fig:30:contracts_cas}
\end{figure}
\fi

\section{Evaluation}
\label{sec:evaluation}

\newcolumntype{Y}{>{\centering\arraybackslash}X}
\newcolumntype{P}[1]{>{\centering\arraybackslash}p{#1}}
\begin{table}[t]
    \caption{The average verification times and number of lines for various examples.}
\begin{tabularx}{\textwidth}{|P{50pt}|P{60pt}|Y|Y| Y | Y | Y |}\hline
    \textbf{Example} & \textbf{Source} & \textbf{\# accesses} & \textbf{Avg. Verif.\ Time (s)} & \textbf{Nr. of lines for protocols} &  \textbf{Nr. of lines for program} & \textbf{Total nr.\ of lines} \\\hline
ARM-weak  & \cite{kang2017promising}                          & 6 & 81.14 & 206 & 133 & 751 \\
COH       & \cite{kang2017promising},\cref{fig:intro_example} & 4 & 55.52 & 100 & 167 & 480 \\
COH-c     & Based on \cite{kang2017promising}                 & 4 & 56.95 & 100 & 161 & 474 \\
COH-V     & \cite{VafeiadisCAV2017}                           & 4 & 49.61 & 100 & 160 & 473 \\
COH-p1    & Based on \cite{VafeiadisCAV2017}                  & 4 & 52.01 & 95  & 160 & 468 \\
COH-p2    & Based on \cite{VafeiadisCAV2017}                  & 4 & 48.30 & 94  & 160 & 467 \\
COH2      & Based on \cite{VafeiadisCAV2017}                  & 6 & 75.52 & 121 & 199 & 533 \\
COH3      & Based on \cite{VafeiadisCAV2017}                  & 9 & 91.68 & 163 & 183 & 559 \\
CoRR2     & \cite{SvendsenPDLV18}							  & 6 & 75.42 & 127 & 143 & 483 \\
dCAS      & \cref{fig:casexample} 					 		  & 2 & 57.16 & 100 & 172 & 619 \\
LB        & \cite{kang2017promising,VafeiadisCAV2017}         & 4 & 69.40 & 143 & 138 & 693 \\
SB        & \cite{kang2017promising,VafeiadisCAV2017}         & 4 & 69.41 & 131 & 130 & 673 \\
2+2W      & \cite{kang2017promising},\cref{fig:intro_example} & 6 & 80.88 & 156 & 143 & 711 \\
    \hline
\end{tabularx}
    \label{tab:42:exprs}
\end{table}	

Thirteen examples have been encoded using the approach, showing different aspects of the encoding. Seven examples have been selected from different papers. Since this work only focuses on the relaxed memory order, all examples with only relaxed memory accesses were picked. Five more examples are based on the examples from the papers. First, COH-c is based on the COH example from \cite{kang2017promising}, reasoning that both threads read 1 instead of 1 and 2 respectively. COH2 and COH3 are also based on the COH example, with more accesses and threads. The examples COH-p1 and COH-p2 are based on COH from \cite{VafeiadisCAV2017}. Both examples show incorrect protocols w.r.t the program, where the first protocol misses a transition and the second protocol specifies the wrong value to write. In all the examples, the type for the values $\Values$ is an integer type. 

\begin{figure}[!t]
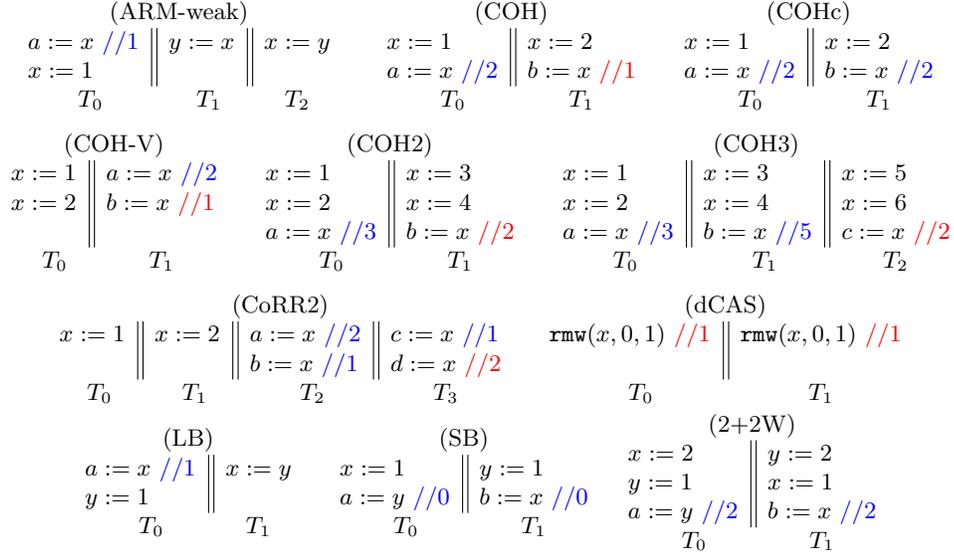

    \centering
    {\begin{tabular}{c} 
\begin{tabular}{l l || l l l || l l }
	\multicolumn{7}{c}{(ARM-weak)}                                              \\                    
	$ a := x\ \textcolor{blue}{// 1} $ & & & $ y := x $       & & &  $ x := y $ \\		            
	$ x := 1 $                         & & &                  & & &  $        $ \\				    
	\multicolumn{2}{c}{$T_0$}&&\multicolumn{2}{c}{$T_1$}&& \multicolumn{1}{c}{$T_2$}\\                    
\end{tabular} $\ \ \ \ $

\begin{tabular}{l l || l l }
	\multicolumn{4}{c}{(COH)}                                                 \\
	$ x := 1 $                         & & & $ x := 2$                        \\
	$ a := x\ \textcolor{blue}{// 2} $ & & & $ b := x\ \textcolor{red}{// 1}$ \\
	\multicolumn{3}{c}{$T_0$}&\multicolumn{1}{c}{$T_1$}                              \\ 
\end{tabular}$\ \ \ \ $
	
\begin{tabular}{l l || l l}
	\multicolumn{4}{c}{(COHc)}                                                  \\
	$ x := 1 $                         & & & $ x := 2$                          \\
	$ a := x\ \textcolor{blue}{// 2} $ & & & $ b := x\ \textcolor{blue}{// 2}$  \\
	\multicolumn{3}{c}{$T_0$}&\multicolumn{1}{c}{$T_1$}                              \\ 
\end{tabular}\\\vspace*{-5pt}\\

\begin{tabular}{l l || l l }
	\multicolumn{4}{c}{(COH-V)}                            \\
	$ x := 1 $ & & & $ a := x \ \textcolor{blue}{// 2}$    \\
	$ x := 2 $ & & & $ b := x\ \textcolor{red}{// 1}$      \\
	$ $ & & & $ $      \\
		\multicolumn{3}{c}{$T_0$}&\multicolumn{1}{c}{$T_1$}                              \\ 
\end{tabular}$\ \ \ \ $

	\begin{tabular}{l l || l l }
	\multicolumn{4}{c}{(COH2)}              \\                                        
	$ x := 1 $ & & & $ x := 3 $                                                     \\
	$ x := 2 $ & & & $ x := 4 $                                                     \\
	$ a := x\ \textcolor{blue}{// 3}  $ & & & $ b := x\ \textcolor{red}{// 2} $     \\
		\multicolumn{3}{c}{$T_0$}&\multicolumn{1}{c}{$T_1$}                              \\ 
\end{tabular}$\ \ \ \ $

\begin{tabular}{l l || l l l || l l}
	\multicolumn{7}{c}{(COH3)}                                                                                           \\
	$ x := 1 $                          & & & $ x := 3 $                        & & & $ x := 5 $                         \\
	$ x := 2 $                          & & & $ x := 4 $                        & & & $ x := 6 $                         \\
	$ a := x\ \textcolor{blue}{// 3}  $ & & & $ b := x\ \textcolor{blue}{// 5} $ & & & $ c := x\ \textcolor{red}{// 2} $ \\
	\multicolumn{2}{c}{$T_0$}&&\multicolumn{2}{c}{$T_1$}&& \multicolumn{1}{c}{$T_2$}\\                    
\end{tabular}\\\vspace*{-5pt}\\

\begin{tabular}{l l || l l l || l l l || l l}
	\multicolumn{10}{c}{(CoRR2)}                                                                                         \\
	$ x := 1 $     & & & $ x := 2 $ & & & $ a := x\ \textcolor{blue}{// 2}   $  & & & $ c := x\ \textcolor{blue}{// 1} $ \\
	$        $     & & & $        $ & & & $ b := x\ \textcolor{blue}{// 1}   $  & & & $ d := x\ \textcolor{red}{// 2}  $ \\
	\multicolumn{2}{c}{$T_0$}&&\multicolumn{2}{c}{$T_1$}&& \multicolumn{2}{c}{$T_2$}&& \multicolumn{1}{c}{$T_3$}\\                    
\end{tabular}$\ \ \ \ $

\begin{tabular}{l l || l l }
	\multicolumn{4}{c}{(dCAS)}                                                                  \\
	$ \casname(x,0,1)\ \textcolor{red}{// 1}$  & & & $ \casname(x,0,1)\ \textcolor{red}{// 1}$ 	\\
	$ $  & & & $ $ 	\\
	
		\multicolumn{3}{c}{$T_0$}&\multicolumn{1}{c}{$T_1$}                              \\ 
\end{tabular}\\\vspace*{-10pt}\\

\begin{tabular}{l l || l l }
	\multicolumn{4}{c}{(LB)}                                    \\
	$ a := x\ \textcolor{blue}{// 1}   $    & & & $ x := y $ 	\\
	$ y := 1 $                              & & & $ $ 			\\
		\multicolumn{3}{c}{$T_0$}&\multicolumn{1}{c}{$T_1$}                              \\ 
\end{tabular}$\ \ \ \ $

	\begin{tabular}{l l || l l }
	\multicolumn{4}{c}{(SB)}                                                     \\
	$ x := 1 $                        & & & $ y := 1 $ 				             \\
	$ a := y\ \textcolor{blue}{// 0}$ & & & $ b := x\ \textcolor{blue}{// 0}$ 	 \\
		\multicolumn{3}{c}{$T_0$}&\multicolumn{1}{c}{$T_1$}                              \\ 
\end{tabular}$\ \ \ $

\begin{tabular}{l l || l l }
	\multicolumn{4}{c}{(2+2W)}                        \\
	$ x := 2 $ & & & $ y := 2 $ 				         \\
	$ y := 1 $ & & & $ x := 1 $ 						 \\
	$ a := y\ \textcolor{blue}{// 2} $ & & & $ b := x\ \textcolor{blue}{// 2} $ 						 \\
		\multicolumn{3}{c}{$T_0$}&\multicolumn{1}{c}{$T_1$}                              \\ 
\end{tabular}
    \end{tabular}}
    \caption{The examples, including their executions}
    \label{fig:evaluation_examples}
\end{figure}

The examples, including the executions that are reasoned about, can be found in \cref{fig:evaluation_examples}. Similar to \cref{fig:intro_example}, the allowed results are shown in blue, while inconsistent results are shown in red. The evaluation verifies that the executions (as described by the comments) are consistent or inconsistent. The cases COH-p1 and COH-p2 use the same program as COH-V, where COH-p1 has a wrong edge for $T_0$ from value 0 to value 2 and COH-p2 has a protocol state with 3 (instead of 2). These cases show that the protocol is also checked against the program.

\Cref{tab:42:exprs} shows the results of VerCors' verification times, a reference to the source of the example, the number
of load/store accesses, and the number of lines in the encoding.\footnote{The encodings of the examples can be found at \cite{spin_weakmem_artifact}} The verification time includes the verification of the protocol against the program and the reasoning about the consistent and inconsistent executions as found in \cref{fig:evaluation_examples}.
The numbers of lines to formalise the protocols, the main program,
excluding protocols and atomic variables,
and the complete encoding, including the atomic variables are given.
The experiments were performed on a MacBook Pro 2020 (macOS 15.7.3) with a 2.0GHz Intel Core i5. Each experiment was performed ten times, after which the average time of those executions was recorded. The time has been measured using the \texttt{time} command.

The verification times for the examples are reasonable, around ~1/1.5 minutes. The examples are of similar size, with between 2 and 9 relaxed loads/stores per example, which explains the similar verification times. The number of lines for four protocol states is around 100, e.g., the COH example of which ~60 lines is boilerplate code for the protocol (including newlines and comments). It grows exponentially with the number of states, due to having to specify that states are distinct, e.g., the ARM-weak protocol. The number of lines of the examples depends on the number of atomic accesses and lemmas to help the prover. The number of lines per atomic variable is common across all examples, which is 203 lines including helper methods and lemmas, except for the dCAS example which includes the \casname method, where the atomic variable encoding requires 333 lines.


\section{Related Work}
\label{sec:10:relwork}
There are numerous memory models that define different semantics for weak atomics and fences. The most notable is the C11/C++11 memory model, which can be translated into hardware memory models. Variants have been made to improve C11, such as RC11~\cite{lahav2017repairing}, RC20~\cite{MargalitRobustnessPOPL21} and the 'promising' semantics~\cite{kang2017promising,promising20}.

The research around weak memory is mainly centered around program logics and semantics that express weak atomics and fences of different weak memory models. These program logics extend and build on top of each other. RSL (Relaxed Separation Logic)~\cite{vafeiadis2013relaxed} is an extension of CSL (Concurrent Separation Logic) to reason about programs under the C11 memory model. FSL (Fenced Separation Logic) and FSL++~\cite{doko2015program,doko2017tackling} extend RSL with fence support. GPS (\textit{Ghost state, Protocols, and Separation})~\cite{2014_GPS_turon}  uses per-location protocols. Basic coherence properties can be established using these protocols~\cite{VafeiadisCAV2017}. GPS+~\cite{he2018gpsp} and GPS++~\cite{2020_gpspp} extend GPS with support for fences.
SLR~\cite{SvendsenPDLV18} is a separation logic for the memory model using a `promising' semantics~\cite{kang2017promising}. SLR integrates the different features from RSL (and extensions) and GPS (and extensions) to tackle programs under weak-memory while avoiding \textit{out-of-thin-air} values. Most of these logics have been formalized in Rocq and their soundness has been proven. Our approach differs from the approach with Rocq mainly in the automization of proofs where the proofs with the logics are manual. 

Weak-memory reasoning by means of static analysis has been considered numerous times, e.g.~\cite{dontsit,stability,Alglave10,effectiveprogramverification,Fang-03,Lee-01,Sura-05}. While this is suitable to approximate which
executions are inconsistent, they do not address the subsequent verification of consistent executions. Our approach allows verifying
that methods adhere to their contracts, considering all consistent executions.

Model checking has also been used to verify programs w.r.t.\ weak memory models~\cite{memorax,Abdulla-15,Atig-11,Bouajjani-13,gpumc,Jonsson,Kuperstein-11,Linden-13,PuWij20}. Compared to deductive verification
approaches, they do not allow modular verification, and are focussed on verifying temporal logic formulae, as opposed to
pre- and postcondition contracts.

There are also works on encoding weak memory reasoning in a deductive verifier. Jacobs encoded the TSO memory model in the VeriFast tool~\cite{jacobs2014verifying}. Summers and Müller~\cite{summers2020automating} encoded large fractions of RSL, FSL and FSL++ in Viper~\cite{MullerSS16}. With Viper, they reason about programs under the C11 model automatically. Both the VeriFast and Viper approaches use a tool that employs separation logic, but they encode different memory models. 
Summers and Müller left the encoding of GPS as future work, which is required to encode coherence properties. 
However, \emph{per-location coherence}, or \emph{SC-per-location}, i.e., the fact that all threads observe the writes to a specific location in a single total order, cannot \textit{explicitly} be \textit{reasoned} about with the logics they support. Our approach is mainly different from this by the fact that it uses view-based protocols to encode coherence properties and support write speculations.

VerCors has support for SC atomics. The work of Amighi~\cite{amighi2014resource,Amighi18Chap5} explains how protocols and state invariants allow for transference. Amighi describes the general contracts for the load, store and read-modify-write operations on SC atomics. The current work extends the support for atomics by encoding the behavior of relaxed memory accesses and introduces per-thread protocols and more granular local views to allow reasoning about coherence properties.


\section{Conclusion}
\label{sec:50:conclu}
We presented an approach to verify weak memory programs using view-based protocols in VerCors-relaxed. The approach allows us to reason about executions of programs, with support for speculative reads using view-based protocols in VerCors-relaxed. 
We discussed how SLR program logic is encoded using the view-based protocols and verified examples using VerCors-relaxed.
Future work involves generating protocol encodings of more examples, since the current protocol generation is tailored towards the set of examples. 
Other future work involves investigating the scalability of this method to larger programs, adding support for release/acquire fences, and targeting other weak memory models.

\subsubsection{Data Availability.} An artifact containing our implementation, our test cases, and scripts that reproduce our experiments and the results presented in \cref{sec:evaluation} is archived and available at~\cite{spin_weakmem_artifact}.

\bibliographystyle{splncs04}
\bibliography{verCorsNK}

\newpage

\appendix

\section{General contract for parallel programs with atomics}
\label{sec:gencontract}

\begin{figure}[!t]
	\begin{lstlisting}
a_V  x = new a_V(N);#\label{lst:30:atomic_init}#
invariant inv( /*@ Global invariant
  Perm(x.G[*], 1\2) ** #\label{lst:30:ginv:pss}# 
  Perm(x.L[*], 1\2) ** #\label{lst:40:ginv:vs}#
  (#\frl# int t1, t2=0..N; intra(x.L[t1][t2]) <= intra(x.G[t2]) || #\label{lst:40:LG:LbehindG}#
						intra(x.G[t2]) <= intra(x.L[t][t2])) @*/) { #\label{lst:40:LG:GbehindL}#
	par (int t=0..N) // Thread contract
	/*@ context Perm(x.G[t], 1\2) ** Perm(x.L[t], 1\2);#\label{lst:30:tlinv:pss}##\label{lst:30:tlinv:vs}#
			context x.G[t] == x.L[t][t]; #\label{lst:30:consistentglobloc}#
			requires x.G[t] == s0(t);#\label{lst:30:initpss}#
			requires (#\frl# int t1; x.L[t][t1] == sx0(t); #\label{lst:30:initvs}#
			ensures isF(t, x.G[t]);#\label{lst:30:acceptingstates}# @*/
	{/* body */}	
}
if (!#\frl# int t1, t2=0..N; intra(x.L[t1][t2]) <= intra(x.G[t2]))
	{/*exclude inconsistent views*/}#\label{lst:30:inconsistentviews}#
	\end{lstlisting}
	\caption{General template for using relaxed atomics on a location \texttt{x}.}
	\label{fig:30:global_local_view}
\end{figure}

With protocols defined (see \cref{fig:30:adt_protocol}) we continue with the encoding of the local views and an (atomic) invariant. \cref{fig:30:global_local_view} shows the general structure of a parallel program with atomics. Note that some parts of the invariant are omitted for readability, such as null checks and arrays lengths. 

The atomic location \texttt{x} (\cref{lst:30:atomic_init}) is an instance of the class \texttt{a\_V}. The class has four methods, namely the constructor, \getname, \setname, and \casname. The contracts of these methods are in \cref{sec:40:encoding}. The class \texttt{a\_V} has two fields \texttt{L} and \texttt{G}. The field \texttt{L} contains the (thread-)local views, which is an array of sequences (i.e., an immutable list) of \texttt{ps} where $\forall {t\in\Tids}. \text{\texttt{\ L[t]}} = \Local_x^{t}$. The encoding introduces a global view \texttt{G}. This view is an array of \texttt{ps}, where $\forall {t\in\Tids} . \texttt{L[t][t]} = \texttt{G[t]}$, i.e., the global view is consistent with every thread's view of its protocol state. \texttt{G} is used as a shorthand to ease verification, with the relation between \texttt{G} and \texttt{L} being maintained throughout the program (e.g., at \cref{lst:30:consistentglobloc}). 

The invariant has two main parts. Firstly, using fractional permissions, we keep half of the permissions to each location of \texttt{G} and \texttt{L} in the invariant \texttt{inv} (\cref{lst:30:ginv:pss,lst:40:ginv:vs}). Keeping half permission in the invariant allows any thread to (atomically) read from the view of another thread. The other half of the permissions is given to the individual threads (\cref{lst:30:tlinv:pss,lst:30:tlinv:vs}). This allows every thread \texttt{t} to update its view \texttt{x.L[t]} when reading or writing. Secondly, the relation between the views of all threads is expressed, where either thread \texttt{t} has speculated about the writing of thread \texttt{t1} (\cref{lst:40:LG:GbehindL}) or not (\cref{lst:40:LG:LbehindG}), encoding condition CC2 mentioned in \cref{sec:20:protocols}.

The contract of threads is structured as follows. The global view is in the initial state (\cref{lst:30:initpss}) and all protocols in the local view are also in their initial states (\cref{lst:30:initvs}). Finally, all threads ensure their global view ends up in an accepting state (\cref{lst:30:acceptingstates}), encoding condition CC1 mentioned in \cref{sec:20:protocols}.

After the parallel block, a case distinction is made (\cref{lst:30:inconsistentviews}). In the end, a view is consistent iff there are no speculations anymore, i.e., the property in \cref{lst:40:LG:LbehindG} holds. We only reason about consistent non-speculating views, otherwise the state is discarded.

\section{Read-Modify-Write}
\label{app:cas}
With the operational semantics for \getname and \setname operations in \cref{sec:20:protocols}, we can define operational semantics for \texttt{\casname} operations. A case distinction is made between a successful and unsuccessful \texttt{\casname} operation.
A successful \texttt{\casname} operation reads the result of some \texttt{\getname} operation, followed by a \texttt{\setname} operation, putting a new constraint on an execution: no other
\casname operation may read the result of the same \setname operation, i.e., atomicity of the \texttt{\casname} operation.

\begin{figure}[!t]
\centering
	\mbox{\inference[(RMW-S)]
		{t, t' \in \Tids \quad
			s \in \Read_x^t(\Local, \Latest, \Time) \cap S_x^{t'} \quad
			\Local_x^t(t) \xrightarrow{w}_{t,x} s' \quad 
			w' = V_x^{t'}(s)
		}{
			\begin{array}{@{}r@{}}
				\langle\Local, \Latest, \Time, \CASsuccess \rangle 
				\xrightarrow{\casname(x, w', w)}_{t,x} 
				\begin{tabular}{l}
				$\langle$\\
				$\ \Local[(t,x) \mapsto \Local_x^t [ t' \mapsto s][t \mapsto s']]$,\\
				$\ \Latest[x \mapsto s']$,\\
				$\ \Time[x \mapsto \Time(x) \cup \{ (\Latest(x), s), (s, s') \}$,\\
				$\ \CASsuccess[s \mapsto \CASsuccess(s){+}1]$\\
				$\rangle$
				\end{tabular}
			\end{array}
		}
	}
\vspace{2ex}

	\mbox{\inference[(RMW-F)]
		{t, t' \in \Tids \quad
			s \in \Read_x^t(\Local) \cap S_x^{t'} \quad
			V_x^{t'}(s) \neq v
		}{
			\langle \Local, \Latest, \Time, \CASsuccess \rangle 
			\xrightarrow{\casname(x, v, \_)} 
			\begin{tabular}{l}
				$\langle$\\
				$\ \Local[(t,x) \mapsto \Local_x^t [ t' \mapsto s]],$\\
				$\ \Latest[x \mapsto s],$\\
				$\ \Time[x \mapsto \Time(x) \cup \{ (\Latest(x), s)\}]$, \\
				$\ \CASsuccess$\\
				$\rangle$
			\end{tabular}
		}
	}
	\caption{Operational semantics for protocol views with successful and unsuccessful \casname operations.}
	\label{fig:SOS2}
\end{figure}

In the operational semantics, this constraint is formalized by keeping track of a \emph{multiset} $\CASsuccess$ of states
from which \texttt{\casname} operations have read values. We interpret this multiset as a function $\CASsuccess: \States \rightarrow \Nat$,
with $\Nat$ the domain of natural numbers.

\cref{fig:SOS2} presents the SOS rules RMW-S and RMW-F, respectively for successful and unsuccessful \texttt{\casname} operations. Essentially, these are combinations of the SOS rules for
\texttt{\setname} and \texttt{\getname}.
The only difference is that the verification states
have been extended with $\CASsuccess$, and after execution of \texttt{\casname\!(x,\!w',\!w)}, $\CASsuccess$ has been updated to
$\CASsuccess[s \mapsto \CASsuccess(s){+}1]$, i.e., $s$ has been added to the multiset $\CASsuccess$ where $V_x^t(s) = w'$. In the rule RMW-F, \CASsuccess is not updated. The Write and Read rules should now also be interpreted with 
\CASsuccess being part of verification states, but it plays no role there.

Once the verification procedure for a program with \casname operations has finished, the final consistency check involves two more conditions.
\begin{equation}
	\forall s \in \States. \CASsuccess(s) = 0 \vee \CASsuccess(s) = 1 \tag{CC3}
\end{equation}
\begin{equation}
	\forall x \in \ALoc, s, s' \in \States_x. (s \neq s' \wedge \CASsuccess(s) =  \CASsuccess(s') = 1) \implies s \not\in \InitStates_x \vee
	s' \not\in \InitStates_x \tag{CC4}
\end{equation}

CC3 expresses that every state occurs at most once in $\CASsuccess$. If a state would occur more than once, this would mean that
multiple \texttt{\casname} operations have read the result of the \setname operation, violating atomicity. CC4 addresses
the special case for initial states: for every $x \in \ALoc$, no two distinct states in \CASsuccess may be both initial, otherwise, again,
atomicity would be violated.

\begin{figure}[t]
	\begin{subfigure}[h]{0.5\textwidth}
		\centering
			\begin{tabular}{l l || l l}
				\casname$(x,0,1)$ & & & \casname$(x,0,1)$\\
				\multicolumn{4}{c}{$T_0$\ \ \ \ \ \ \ $T_1$}\\
			\end{tabular}
	\end{subfigure}
	\begin{subfigure}[h]{0.4\textwidth}
			\includegraphics{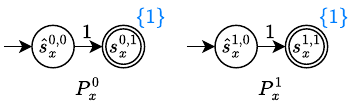}
	\end{subfigure}
	\caption{An example program with \casname operations (left) and its protocols (right).}
	\label{fig:casexample}
\end{figure}

\begin{example}
	Consider the example program and its protocols in \cref{fig:casexample}.
	Since both \casname operations expect to read the same value, and since they both write a value different from that value in case of success, exactly one of the instructions should succeed. 
	Initially, we have $\Local_x^0(0) = \Local_x^1(0) = \stti{x}{0}{0}$ and $\Local_x^0(1) = \Local_x^1(1) = \stti{x}{1}{0}$. 
	First, for $T_0$, $\casname(x,0,1)$ is can succeed, as $\Read_x^0(\Local) = \{ \stti{x}{0}{0}, \stti{x}{1}{0}, \stt{x}{1}{1} \}$. 
	For this, we can either select $\stti{x}{0}{0}$ or $\stti{x}{1}{0}$, update $\Latest$ and $\Time$ accordingly, and add one of those states to \CASsuccess. Let us add $\stti{x}{0}{0}$ to \CASsuccess, the case for $\stti{x}{1}{0}$ is similar. 
	Next, if we consider the \casname instruction of $T_1$, we have to conclude that only the failing $\casname(x,0,1)$ is possible; for a \texttt{\casname} operation, we would again have to select either $\stti{x}{0}{0}$ or $\stti{x}{1}{0}$. 
	In the case that we add $\stti{x}{0}{0}$ to \CASsuccess, CC3 is violated, and in the case that we add $\stti{x}{1}{0}$ to \CASsuccess, CC4 is violated. 
	Symmetrically, $T_0$ can execute a failing $\casname(x,0,1)$, in which case $T_1$ can execute a successful $\casname(x,0,1)$. 
	The case that neither \casname instructions succeeds is ruled out as it would involve invalid speculations: 
	if $T_0$'s $\casname(0,x,0)$ operation fails, then after the execution, we must have $\Local_x^0(1) = \stt{x}{1}{1}$. Similarly, for $T_1$, execution of $\casname(x,0,1)$ leads to $\Local_x^1(0) = \stt{x}{0}{1}$. However, in that case, CC2 is violated.
\end{example}

\subsection{Encoding into VerCors}
\cref{fig:30:contracts_cas} shows the contract for the \casname method. The contract of the \casname operation is a combination of the contracts for \getname and \setname. With the contracts for the abstract \getname and \setname methods, the implementation of the \casname method can be verified. For brevity, parts of the \casname's contract overlapping with the contracts of \getname and \setname are omitted to focus on the differences. The contract mainly consists of two parts:

\begin{compactenum}
	\item \textbf{\casname fail}: When the operation fails, the global view of \texttt{t} does not change (\cref{lst:30:cas:nochange}). The call to \getname updates the local view. The return value of \getname cannot be equivalent to \texttt{oldVal} (\cref{lst:30:cas:newval_no_oldval}).
	Finally, a description of \texttt{newLV} can be given (\cref{lst:30:cas:descriptionfail}). The predicate \texttt{descriptionFail} is a placeholder and is to be defined by the user to use in further proofs. It can be used to exclude certain executions, see \cref{fig:casexample}.
	
	\item \textbf{\casname success}: The \texttt{newLV} is specified to have changed similarly to the \setname method's contract. In addition to this, a token \texttt{CASSuccess} is given to signify that the operation has succeeded (\cref{lst:30:cas:cassucces}). This token captures the state that \texttt{oldVal} is read from. The token is used to exclude multiple operations reading from the same write.
	Again, a description of \texttt{newLV}  can be given (\cref{lst:30:cas:descriptionsucceed}) with the predicate \texttt{descriptionSuccess}.
\end{compactenum}

\begin{figure}[!t]
	\begin{lstlisting}		
/*@ given seq<ps> oldLV; 
	yields seq<ps> newLV;
	given int N, tid, depTid, prev;
	yields int next; 
#\textcolor{red}{1)}#ensures !\result ==> G[tid] == \old(G[tid]);#\label{lst:30:cas:nochange}#
	ensures !\result ==> next != oldVal; #\label{lst:30:cas:newval_no_oldval}#
	ensures !\result ==> descriptionFail(newLV);#\label{lst:30:cas:descriptionfail}#
#\textcolor{red}{2)}#ensures \result == G[tid] == ps.next(oldLV[tid]);
	ensures \result ==> newLV[tid] == ps.next(oldLV[tid]);
	ensures \result ==> ps.val(newLV[tid]) == newVal;
	ensures \result ==> next == newVal;
	ensures \result ==> next \in Read(newLV);
	ensures \result ==> CASSuccess(tid,newLV[depTid]);#\label{lst:30:cas:cassucces}#
	ensures \result ==> descriptionSuccess(newLV);#\label{lst:30:cas:descriptionsucceed}#@*/
boolean rmw(int oldVal, int newVal) {#\label{lst:30:a_V:cas}#
	int c = get() given {...} yields {...};
	if (c == oldVal) { return true; } 
	return false; 
}
	\end{lstlisting}
	\caption{The contract of the \casname method.}
	\label{fig:30:contracts_cas}
\end{figure}

\section{The encoding of the COH example}
\cref{app:COH:main} shows the PVL encoding of the COH example from \cref{fig:intro_example}. The encoding of the atomics and some helper methods are omitted for readability.

\begin{figure}
	\centering
	\begin{lstlisting}[basicstyle=\ttfamily\notsotiny]
given int N; context_everywhere NBound(N);
void main() {
	GSX gx = new GSX(N); assume x() == gx; 
	
	invariant inv(/*@
	  NBound(N) ** gspsx(x(), N) **
	  // For x
	  (Perm(x().G[*], 1\2)) ** 
      
	  (#\frl# int t=0..N; ps.N(x().G[t]) == N) ** 
	  (#\frl# int t=0..N; padhereX(x().G[t], t, N)) **
	
	  x().L != null ** 
	  x().L.length == N ** 
	  (#\frl#* int t=0..N; Perm(x().L[t], 1\2)) ** 
	  (#\frl# int t=0..N; |x().L[t]| == N) **
		
	  // G[t] reachable from L[t][t'] or vice versa 
	  (#\frl# int t=0..N, int t1=0..N; 
    	  // L[t] older value than G[t]                         or L[t] speculates into a future of G[t]
    	  ps.intra(x().L[t][t1]) <= ps.intra(x().G[t1]) || ps.intra(x().L[t][t1]) >= ps.intra(x().G[t1])
	  ) ** 
		
	  // G[t] and L[t][t] always consisntent
	  (#\frl# int t=0..N; x().L[t][t] == x().G[t]) @*/) {
		par (int tid=0..N) 
		/*@ context_everywhere NBound(N) ** gspsx(x(), N);
			// For x
			context_everywhere Perm(x().L[tid], 1\2) ** |x().L[tid]| == N;
			requires x().L[tid] == psInitXSeq(N);
			context_everywhere handle(Px(), tid, x().L[tid]) ** handle(Cx(), tid, x().L[tid]);
			context_everywhere Perm(x().G[tid], 1\2);
			context_everywhere x().G[tid] == x().L[tid][tid];
			context_everywhere (#\frl# int t=0..N; padhereX(x().L[tid][t], t, N));
		
			requires x().G[tid] == ps.sx0(tid,N);			
			ensures ps.isF(tid, x().G[tid]); @*/
		{
			int lvx=\choose(ReadX(x().L[tid])); 
			lemmaReadXHelperInit(x().L[tid],0,N); 
			assert lvx == 0; // Ghost state, initial l(ocal)v(iew)
			int wx=tid;
			ps lsx=x().L[tid][wx];
			
			if (tid == 0) {
				atomic(inv) {seq<ps> tmp1; 
					x().store(1) given {oldLV=x().L[tid],N=N,tid=tid,prev=lvx,prevW=wx,prevPi=lsx} 
                         yields {lvx=next,wx=nextW,lsx=nextPi,tmp1=newLV};
					x().L[tid] = tmp1;}
				
				
				int a;
				atomic(inv) {seq<ps> tmp1; 
					a = x().load() given {oldLV=x().L[tid],N=N,tid=tid,prev=lvx,prevW=wx,prevPi=lsx} 
                           yields {lvx=next,wx=nextW,lsx=nextPi,tmp1=newLV};
					x().L[tid] = tmp1;}	
				
			} else {
				atomic(inv) {seq<ps> tmp1; 
					x().store(2) given {oldLV=x().L[tid],N=N,tid=tid,prev=lvx,prevW=wx,prevPi=lsx} 
                         yields {lvx=next,wx=nextW,lsx=nextPi,tmp1=newLV};
					x().L[tid] = tmp1;}
				
				int b;
				atomic(inv) {seq<ps> tmp1; 
					b = x().load() given {oldLV=x().L[tid],N=N,tid=tid,prev=lvx,prevW=wx,prevPi=lsx} 
                           yields {lvx=next,wx=nextW,lsx=nextPi,tmp1=newLV};
					x().L[tid] = tmp1;}
			}
		} 
	}
    // Exclude all inconsistent view, i.e., views with unkept promises.
	if (!x().consistentL()) {inhale false;} 
	refute false;
}
	\end{lstlisting}
	\caption{The encoding of the COH example.}
	\label{app:COH:main}
\end{figure}

\end{document}